\documentclass[aps,pra,showpacs,twoside,twocolumn,longbibliography,10pt]{revtex4-1}
\usepackage[colorlinks=true, citecolor=red, urlcolor=blue ]{hyperref}
\usepackage{epsfig,newlfont,amssymb,amsfonts,amsmath,bm,subfigure,palatino,mathtools,amsthm,braket,times,soul,enumitem,color}
\usepackage[normalem]{ulem}
\newcommand{\stkout}[1]{\ifmmode\text{\sout{\ensuremath{#1}}}\else\sout{#1}\fi}
\usepackage[english]{babel}
\usepackage[utf8]{inputenc}
\usepackage{array}
\usepackage{xcolor}
\usepackage{graphics}

\newcommand{\ketbra}[2]{|#1\rangle \langle #2|}
\def\Tr{\text{Tr}}

\usepackage{amsthm}
\usepackage{verbatim}
\usepackage{bbm}
\usepackage{wrapfig}

\usepackage{hyphenat}

\newlength\figureheight 
\newlength\figurewidth

\begin{document}

\title{Quantum transistors for heat flux in and out of working substance parts:\\harmonic vs transmon and  Kerr environs}



\author{Deepika Bhargava$^{1,2,3}$}
\thanks{These authors contributed equally to this work.}

\author{Paranjoy Chaki$^{2,3}$}
\thanks{These authors contributed equally to this work.}

\author{Aparajita Bhattacharyya$^{2,3}$}
\thanks{Corresponding author: \href{mailto:aparajita96.phy@gmail.com}{aparajita96.phy@gmail.com}}

\author{Ujjwal Sen$^{2,3}$}

\affiliation{$^{1}$Indian Institute of Science Education and Research, Homi Bhabha Rd, Pashan, Pune 411 008, India\\
$^{2}$Harish-Chandra Research Institute,  
Chhatnag Road, Jhunsi, Prayagraj 211019, India\\
\(^3\) Homi Bhabha National Institute,  Training School Complex, Anushakti Nagar, Mumbai 400 094, India
}


\begin{abstract}

Quantum thermal transistors have been widely studied in the context of three-qubit systems, where each qubit interacts separately with a Markovian harmonic bath. In contrast, non-Markovianity is a general feature, inherent to a large fraction of realistic scenarios. Instead of Markovian environments, here we propose a transistor in which the interaction between the working substance and an environment comprising an infinite chain of qutrits is based on periodic collisions. We refer to the device as a working-substance thermal transistor, where the definition of the heat current is considered to be system-centric. We find that the transistor effect also prevails in this scenario. We consider the variation in amplification with respect to the temperature of the modulating bath, the system-environment coupling, and the interaction time. We also investigate how varying the interaction strengths between the terminals affects amplification. Additionally, the environment, comprising three-level systems, allows us to consider the effects of frail perturbations in the energy spacings of the qutrit, leading to non-linearity in the environment. We consider non-linearities that are either of transmon or of Kerr-type. We identify parameter regimes in which transmon and Kerr-type nonlinear environments provide a significant enhancement compared to linear environments. 
\end{abstract}

\maketitle
\section{Introduction}

Advancements in science and technology have led to the development of modern equipment that requires miniaturized electrical circuits. These small-scale components exhibit thermal and quantum properties that impact performance of the device, making control over electrical and thermal fluctuations crucial.
A transistor is a device that controls current flow in electric circuits~\cite{transistor,trans-book-1,trans-book-2,trans-book-3}.
Electrical transistors are essential for modern technology, utilized in amplifiers, switches, and transformers. Since their invention by Bardeen and Brattain~\cite{transistor}, extensive research has focused on improving and adapting transistor-based devices~\cite{trans-book-1,trans-book-2,trans-book-3,transistor,trans-work-old2,trans-work-old3}.

A quantum thermal transistor (QTT)~\cite{qtt-first-paper} is the quantum analog of an electric transistor, used to control heat currents in miniaturized circuits. Quantum effects enable operations not possible in classical systems, offering advantages over traditional devices.
Other quantum thermal machines like quantum batteries~\cite{battery-1,battery-2,battery-3}, quantum refrigerators~\cite{refrigerator-1, refrigerator-2, refrigerator-3}, quantum thermal diodes~\cite{diode}, quantumn heat engines~\cite{KosloffLevy2014,UzdinLevyKosloff2015,NimmrichterRouletScarani2018,Saha2023Harnessing} and quantum rectifiers~\cite{rectifier-1, rectifier-2, rectifier-3} have been devised which exploit quantum mechanical properties of the system. Since the introduction of the quantum thermal transistor by Joulain et al. in 2016~\cite{qtt-first-paper}, significant progress has been made, particularly in systems involving Markovian environments~\cite{gen-1, gen-2, gen-3, gen-4, gen-5, gen-6, gen-7}.



Current QTT models typically utilize a three-qubit working substance, each connected to a Markovian bath. Other models include qubit-qutrit systems~\cite{qubit-qutrit} and those employing periodic control, such as the Floquet quantum thermal transistor~\cite{floquet_tt,Fq_transistor_2}.
There are extensive works
which consider Coulomb-coupled conductors~\cite{columb-coupled-1, columb-coupled-2}, polarons 
in the context of non-equilibrium systems~\cite{polaron}, double cavity optomechanical systems~\cite{cavity-optomechanical}, and three terminal normal-superconductors~\cite{normal-superconductor} as models for QTTs. 
QTT performance, including environmental effects with and without noise, has been extensively studied in single and multi-transistor systems with shared baths.~\cite{noise-1, noise-2, noise-3, noise-4}. 
The significance of quantum coherence in the context of QTT has also been studied, which enables to 
provide quantum advantage in the amplification of a QTT
over its classical counterpart~\cite{coherence-enhanced}. 
Further studies on QTT focus on negative differential thermal conductance (NDTC) for high amplification rates and the role of strong coupling and quantum coherence in achieving NDTC~\cite{ndtc-1, ndtc-2, ndtc-3, ndtc-4}.
%
Models analogous to these classical ones, but employing QTTs, have also been rigorously investigated. In particular, configurations such as the Darlington setup~\cite{darlington-1, darlington-energy-divider}, where the output of one transistor serves as the input to another, or more complex transistor networks as considered in~\cite{transistor-network}, have been shown to provide significant performance enhancements compared to scenarios that do not incorporate QTTs.
Experimental realization of 
QTTs using superconducting circuits~\cite{exp-super-1,exp-super-2}, materials like $VO_2$~\cite{exp-vo2, exp-vo2-2}, 
and implementation using magnetic thermal transistor~\cite{exp_magnetic} have been carried out. 

Previous QTT studies utilize a global heat current approach, treating the system as three qubits where 
the focus is on the
heat flowing from each bath into the entire working substance.
The heat current is "global" in a sense that the heat current corresponding to each of the baths flow globally into the entire working substance.
This paper focuses on local heat currents, where we concentrate on the heat currents flowing ``locally" into each qubit from different parts of the system and environment, in contrast to the global heat current approach. 
We consider the heat flowing in and out of each particular qubit, rather than the heat flowing in and out of a particular environment. Within this framework, each qubit effectively plays the role of a terminal. Here, ``local" refers to each qubit being coupled to a distinct environment.
So the definition of heat current in earlier papers was bath-centric, while in our case it is system-centric.
Our definition is motivated from a realistic fact that a precise experimental determination of the heat current originating from an environment, modeled as a large infinite
bath, is 
highly difficult to implement in physical scenarios. Such a technically challenging procedure is not required in our protocol. Instead, the local heat current can be inferred from observables associated with the individual sub-components of the working substance, which are experimentally accessible.
From an experimental perspective, the properties of the individual sub-parts of the working substance are accessible, and controlling and regulating heat flow at the level of these sub-parts is important for practical implementations in quantum circuits, as well as for preserving the integrity of the material components. This is where the significance of the protocol and the definition of the local heat current become apparent.
Our concept of such local heat current is novel, and is hitherto unexplored in literature.
There have been previous works which 
discuss the
different types of heat currents in thermal equilibrium~\cite{heat-current-1}, where the system is connected to an Ohmic bath with a given spectral density. Here, they have used certain approximations to derive the local master equation and subsequently the definition of the local heat current. These assumptions are as follows.
Firstly, the interaction between system and environment is perturbatively very small. Secondly, the Markov approximation assumption is made  which tells about the invariance of system's state on time scales of order correlation time.
However, these assumptions are not valid in our case. We have defined our system for arbitrary coupling instead of weak coupling which does not follow the Born approximation. The Markov approximation also talks about the bath retaining its state during and after the evolution, which is not true in the dynamics we have considered. Our definition of a local heat current is not equivalent to their definition since the Born Markov assumptions on which they have defined the local heat current are not valid in our system. 

In our case, the working substance is considered under the most general completely positive trace-preserving (CPTP) dynamics, i.e. its evolution is described by a global unitary acting on the joint system of the working substance and its environment, followed by discarding the environment. Importantly, no assumption of Markovianity is made in the analysis, and therefore, Markovian master equations do not naturally arise in this framework.

In Ref.~\cite{heat-current-2}, for instance, it 
has been rigorously proven how local master equations are consistent with the laws of thermodynamics.
As Markovian nature of a thermal environment, in general, arises under some specific assumptions, and is very unlikely to obtain in a practical scenario, it is important to study the performance of quantum transistors in the presence of non-Markovian environments. Quantum thermal devices in the context of non-Markovian environments have been previously delved into~\cite{breuer_,samyadeb,apa_spin_ref},which mainly 
focused on spin environments. 
{There exist other works in the literature employing non-Markovian environments in the context of quantum thermal devices, such as quantum heat engines~\cite{Uzdin2016,Kato2016,Chen2016,Raja2021,Cavaliere2022,Ptaszynski2022, Vacchini2018NonMarkovianThermo},
specifically Otto engines~\cite{Camati2020,DasMukherjee2020,Shirai2021,Chakraborty2022,PhysRevResearch.5.023066}, quantum batteries~\cite{PhysRevB.104.245418,4hr3-rl2y}, spin star systems~\cite{PhysRevB.70.045323,PhysRevA.81.062353} etc.}



However, here we consider the system-environment interaction to be modeled by a collisional model comprising an infinite array of qutrits, which is experimentally feasible to implement and is fundamentally different from the Markovian environment scenario.

We 
provide
a prototype of a QTT, which consists of three qubits interacting with each other and each  qubit is connected separately to an environment. 
The interaction between the system and environment mimics the collisional model~\cite{coll-model-1, coll-model-2, coll-model-3, coll-model-4, coll-model-5},
which has been extensively studied in literature in the context of quantum batteries~\cite{sen2023noisyquantumbatteries, coll-model-battery2}, equilibrium and non-equilibrium dynamics~\cite{coll-model-2, coll-model-3, coll-model-4, coll-model-equi1, coll-model-equi2}, strong coupling thermodynamics~\cite{coll-model-str-coupling} and the impact of quantum coherence in thermodynamics~\cite{coll-model-thermo1, coll-model-thermo2} and others~\cite{c_1,c_2,cc3,C_4,C_5,coll-model-thermo1}.
In 
the
model that we consider, each system qubit sequentially interacts with a qutrit of the environment for a short but fixed duration of time. 
The environment, which originally consists of infinite number three-level systems with equispaced energy levels, also allows us to consider the effects of frail perturbations in the energies of the environment-particles, leading to unequal energy spacings of each qutrit. 
Considering this model as 
the environment instead of the usual Markovian bosonic one, introduces an inherent non-Markovianity in the system.
Using this model, we devise our transistor, which can control and regulate heat currents in the sub-parts of the working substance.

In this work, we scrutinize a class of transistors comprising a three-qubit working substance which interacts with the environment via the collisional model.
There also exist interactions between the components of the working substance, i.e. the left, middle, and right qubits. Initially, we consider equal interaction strength between the left-middle and right-middle qubits, but the left and right qubits remain non-interacting. 
We find significant amplification of the heat currents at two terminals of the transistor, being  modulated by the heat current at the remaining terminal. The amplification factor depends on the temperature of the modulating environment, and system-environment coupling. 
Specifically, the amplification factors for the left and right qubits first increase and then decrease, as the modulating temperature rises.
The amplification trend with coupling strength is complex: it decreases below a critical temperature but increases beyond it with stronger interactions. We also observe some
amplification trend with the temperature of modulating bath
{The observed amplification remains robust over a broad range of system-bath interaction strengths, collision times, numbers of interactions, and initial-state preparations.}
Additionally, in a linear system with equispaced energy levels, external perturbations can induce small fluctuations, effectively rendering the system nonlinear.
Two examples of where such a situation naturally arises are  transmons~\cite{tr1,tr2,tr3,tr_exp,apa} and Kerr-type environments~\cite{kerr1,Avijit_Mishra}.
In our work, apart from considering environment, which constitutes the collisional model of  qubits interacting sequentially with equispaced qutrits, we also
consider the effects of non-linearities in the environment such as in transmon and Kerr-types, in the amplification of a type of transistor.
We find that amplification and its temperature dependence persist even with small non-linearities in the qutrit energy levels. Transmon and Kerr-type non-linearities enhance amplification in different temperature regimes. Below a critical temperature, Kerr non-linearities outperform linear qutrits, while above it, transmon qutrits are more effective. Furthermore, we find a non-monotonic behavior in the amplification factor of the three-qubit transistor setup with respect to time.
{Moreover, as the dynamical profile of the amplification exhibits a sequence of pronounced peaks, we additionally investigate the time-averaged amplification, $(\bar{\alpha}_L)$, to characterize the overall transistor behavior. The averaged amplification likewise exhibits a clear transistor-like response for all three types of environments considered in this work.}

Then, we consider two cases, i.e. symmetric and asymmetric coupling, where for symmetric coupling, left-middle, middle-left, and left-right qubits of working substance are coupled with each other by equal interaction strength. On the other hand, for asymmetric coupling, left-middle, middle-left, and left-right qubits are coupled by different interaction strengths. We show that appreciable amplification is obtainable for both the cases corresponding to both left and right qubits. Moreover, we also observe the benefits of nonlinear environments in obtaining higher amplification in both symmetric and asymmetric cases.

The paper is organized as follows. In Sec.~\ref{prelims}, we 
explicitly describe 
the collisional model for the system and environment that we consider,
along with defining the relevant Hamiltonians, the dynamical amplification factor, and the different types of heat-currents. 
We then analyze the effect of varying temperature of the modulating bath, and coupling constant between the bath and the working substance on the amplification, along with the symmetric and asymmetric cases, considering linear baths in Sec.~\ref{linear-sec}.
We consider both the cases where there are three (Sec.~\ref{sub1}) and less than three (Sec.~\ref{sub2}) environments.
We then introduce non-linearity in the baths and analyze its effect on amplification in Sec.~\ref{non-linear-sec}. 
%
Finally the concluding remarks are presented in Sec.~\ref{conclusions-sec}. 

\begin{figure*}[htbp]
    \subfigure[]{\includegraphics[width=0.4\textwidth]{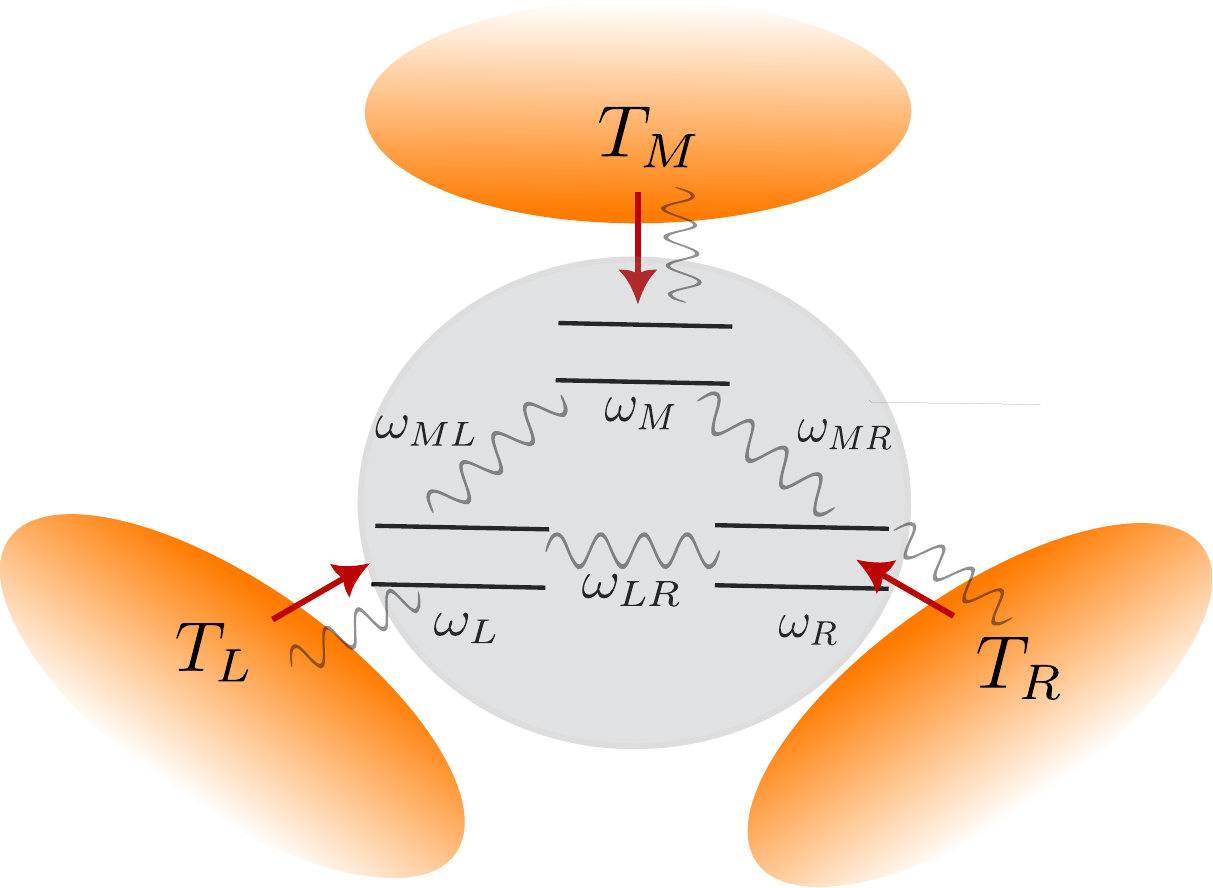}} 
    \label{a}
    \subfigure[]{\includegraphics[width=0.4\textwidth]{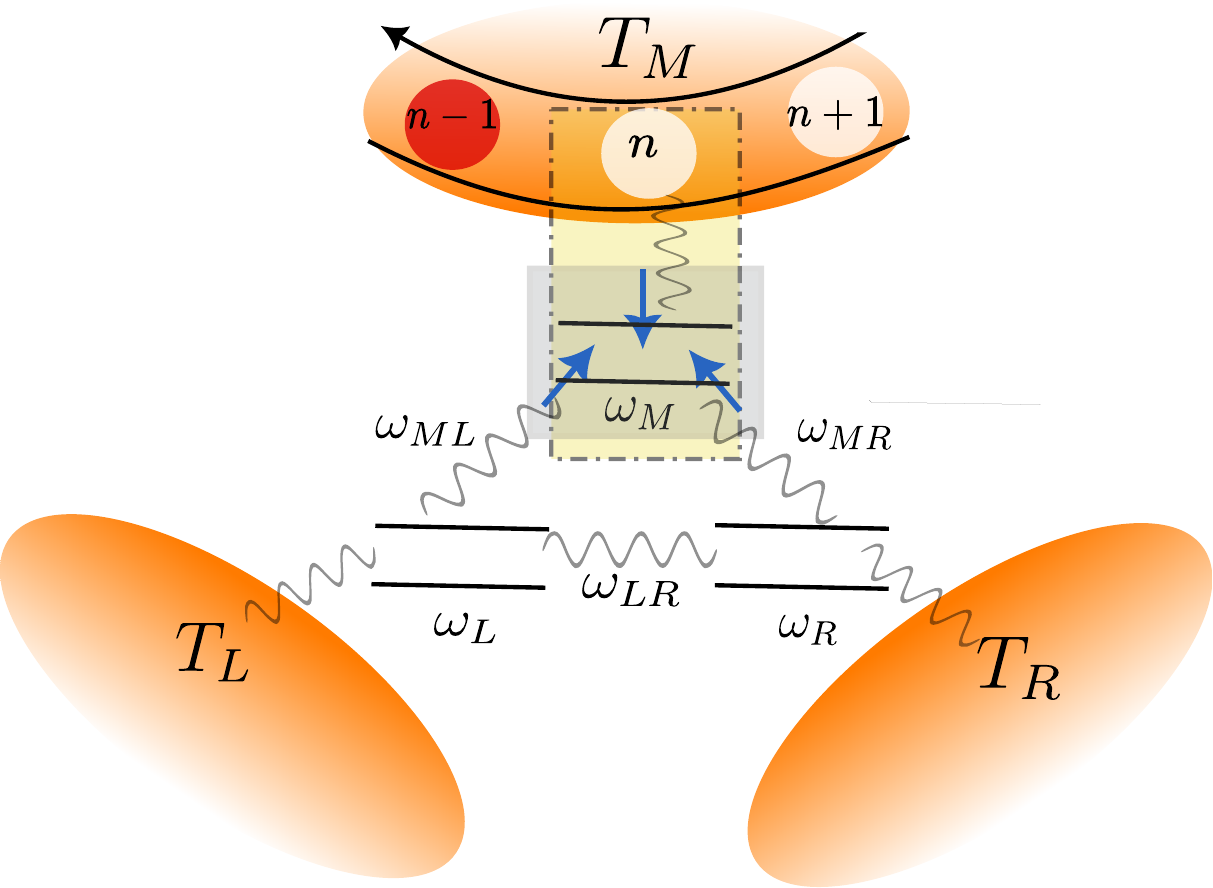}} 
    \label{b}
    \caption{Schematic representation of the difference between the local and global approaches 
    utilized to provide the definition of heat current used in previous models, referred to as BTTs 
    versus the model studied in this paper, referred to as WTTs. 
    The name, BTT essentially means that the transistor is conceptualized by looking at heat current flows in and out of each of the local baths to and from the entire system, as depicted by the red arrows in panel~(a), and hence the transistor action is scrutinized by focusing on the
    ``bath". The name, WTT suggests that, the heat 
    is exchanged between each (local) qubit with the different parts of the system and the environment,
    as shown by the blue arrows in panel~(b), and therefore 
    the focus is on the 
    ``working substance". BTTs correspond to the usual QTT model previously studied in literature, while the class of transistor devices which we analyze in this paper are the WTTs. The bath interacts with the working substance via the collisional model where every qutrit sequentially interacts with the working substance. We show in the schematic that the $n^{\text{th}}$ qubit is interacting with the system while the $(n-1)^{\text{th}}$ qubit has already interacted. The interaction with the $(n+1)^{\text{th}}$ qubit is going to occur. }
    \label{system}
\end{figure*}
\section{Working-substance thermal transistor}
\label{prelims}
An electrical transistor is a three terminal device which consists of a p-type (n-type) semiconductor sandwiched between two n-type (p-type) semiconductors to form a n-p-n (p-n-p) junction. The three terminals, corresponding to n-p-n, are often referred to as emitter, base and collector, and the currents flowing into each of these terminals are the emitter, base and collector currents, denoted by
$I_E, I_B$ and $I_C$ respectively. 
The function of an electrical transistor is to amplify the input base current to the output collector current, while operating in the common emitter mode.
The figure of merit that quantifies the performance of a transistor is given by the dimensionless quantity, direct current (DC) amplification factor, $\beta_{DC}$, which is defined as
\begin{equation}\nonumber
    \beta_{DC} = \frac{I_C}{I_B}.
\end{equation}
This figure of merit is useful, particularly in transistor circuits where all the currents are constant with respect to time.
Meanwhile, if the currents are functions of time, the alternating current (AC) amplification factor, $\beta_{AC}$, is defined as 
\begin{equation}\nonumber
    \beta_{AC} = \frac{\Delta I_C}{\Delta I_B},
\end{equation}
where $\Delta I_C$ and $\Delta I_B$ are the small changes in $I_C$ and $I_B$ in the AC circuit of the transistor. The amplification factor refers to the change in collector current due to a small change in base current. Typically, the values of $\beta_{DC}$ or $\beta_{AC}$ is of the 
order of 200-300 and is an intrinsic 
property of an electrical transistor.

A QTT, on the other hand, is a thermal device which amplifies and modulates heat currents flowing into it, 
in analogy to an electrical transistor which amplifies electric currents. 
The quantum thermal transistor that we consider here consists of three two-level systems, which we refer to as the working substance. Each of these two-level systems is coupled separately to thermal environments. The interaction between the system and environment mimics that of the collisional model which we define in the succeeding subsection.
We refer to the three two-level systems as $L, M,$ and $R$ which correspond to the left, middle, and right qubits, respectively.
The energy differences between the two levels of the two-level systems are, respectively, given by $\hbar\omega_L, \hbar\omega_M$ and $\hbar\omega_R$. The  temperatures of the three thermal environments attached to the qubits are $T_L, T_M$ and $T_R$ respectively. The middle environment is taken to be the modulating one which modulates the current in the left and right qubits analogous to its electrical counterpart. 
The dynamical amplification factor, $\alpha_X$ corresponding to the qubit-$X$, of a QTT is defined as
\begin{equation}
    \alpha_X = \frac{\partial J_X}{\partial J_M} = \frac{\frac{\partial J_X}{\partial T_M}}{\frac{\partial J_M}{\partial T_M}}
    \label{amp-factor-eqn}
\end{equation}
where $X \in \{L,R\}$. The quantity, $J_X$, denotes the heat current flowing into the
terminal $X$ of the
working substance. 
A more elaborate discussion on heat currents, focusing on the type of heat current that we consider, is explicitly provided in Sec.~\ref{heat-current-def-sect}.
The magnitude of amplification factor is indicative of how well the transistor performs, with the value of $|\alpha_X|$ being greater than $1$ indicating that the transistor effect exists. 


\subsection{Collisional model for system-environment interaction: linear and non-linear}
The transistor models, that have been hitherto explored, fundamentally constitute local Markovian environments connected to each qubit of a QTT, where the notion of heat current corresponding to the individual heat flow from each bath was relevant. Here we use a different approach to design a type of transistor model, and provide a somewhat distinct definition of heat current, where we look into the heat current flowing into each of the qubits separately. This is discussed in detail in subsection~\ref{heat-current-def-sect}. Here we focus on the model of the environment that we consider.
The working substance comprises three qubits, where each qubit is open to an environment. The environment constitutes an infinite number of three-level systems or qutrits. Each qutrit  arrives one after the other, and interacts with a single qubit of the working substance for a fixed time, say $\delta t$. After each time interval, $\delta t$, a new qutrit pertaining to the environment arrives to interact with the qubit. Hence, the environment 
effectively resets after the finite interaction time, $\delta t$. 
Owing to the periodic collisions of the system with the environment, this model is often referred to in literature as the ``collisional model"~\cite{coll-model-1,coll-model-2,coll-model-3,coll-model-4,coll-model-5}.

Initially we consider the energy difference between the ground and first excited states of each  qutrit, that comprises the environment, to be equal to that between the first and second excited states of the qutrit. Since both the energy levels are equispaced, we refer to the type of environment comprising such qutrits to be a linear one. 
%
%
However in reality, there may be frail perturbations in the environment leading to small differences in the energy-level spacings.
This introduces non-linearity in the environment due to a tweaking of the energy levels by a small amount.
Let us say that the energy of the first excited level is increased by a small quantity. 
Then the energy gap between the ground state and the first excited state is greater than the  difference between the first excited state and the second excited state.
If the energy level differences decrease as we go to higher energy levels,
then we get a transmon environment. 
Whereas if
the energy level differences increase as we go from first to second excited states,
a Kerr-type environment is generated. We refer to 
both these types of environments as non-linear ones.

\subsection{Relevant Hamiltonians}\label{BBBBBBBBBBB}
In this subsection, we discuss the transistor model that we consider, including both the system and environment Hamiltonians, and the interactions between them.
The Hamiltonian of the system comprising the three qubits is given by

\begin{equation}
    H_{sys} =  - \sum_{i\in\{L,M,R\}} \frac{\hbar \omega_i}{2}~\sigma_{z}^i -  \sum_{\substack{i,j \in\{L,M,R\} \\ i\neq j}} \frac{\hbar \omega_{ij}}{2}~\sigma_z^i~\sigma_z^j,
\label{sys_hamiltonian}
\end{equation}
where $\sigma_z$ refers to the Pauli-$z$ matrix for spin-$1/2$ particles.
Here $L,M$ and $R$ refers to the left, middle and right qubits respectively.
The first term represents the local Hamiltonian of each qubit, while the second term denotes the interaction between the qubits. 
{We assume $\omega_i,\omega_{ij}>0$, which fixes the level ordering of the Hamiltonian. Consequently, the ground state of the system is
$
|G\rangle = |000\rangle_{q},
$
where $\ket{0}_{q}$ is the eigenstate of $\sigma_z$ with eigenvalue $+1$.}
The local Hamiltonian of a single qutrit of the environment is given by

\begin{equation}
    H_{env} = -\hbar~ \Delta \sum_{i\in\{L,M,R\}} \sigma_z^{(i),env},
\label{bath-eqn}
\end{equation}
where the energy levels of each qutrit are equispaced, and $\Delta$ is the  energy gap between any two successive energy levels. Here $\sigma_z^{(i),env}$ is the Pauli-$z$ matrix for spin-$1$ particles. The index $i$ denotes the qutrit connected to $i^{\text{th}}$ qubit of the system.
The Pauli-$x$ and Pauli-$z$ matrices for spin-$1$ are respectively given by

\begin{equation}
\nonumber
 \sigma_x^{env} = \frac{1}{\sqrt{2}} \begin{bmatrix}
0 & 1 & 0\\
1 & 0 & 1\\
0 & 1 & 0\\
\end{bmatrix}, 
    \sigma_z^{env} =  \begin{bmatrix}
1 & 0 & 0\\
0 & 0 & 0\\
0 & 0 & -1\\
\end{bmatrix}.
\end{equation}
{Assuming $\Delta>0$, the ground state of $H_{env}$ is
$
|G\rangle_{env} = |000\rangle_{env},
$
where $\ket{0}$ is the eigenstate of $\sigma^{env}_z$ with eigenvalue $+1$.}
In Sec.~\ref{linear-sec}, we consider the collisional model comprising of equispaced qutrits, which arrive one at a time to interact with the system qubit.
Each of the three qubits interact locally with each qutrit of the environment via the qubit-qutrit interaction given by 

\begin{equation}
    H_{qubit-env} = - \hbar~g\sum_{i\in\{L,M,R\}}\sigma_x^{(i),qubit}~\sigma_x^{(i),env}
    \label{qubit-env-eqn}
\end{equation}
where $\sigma_x^{(i),qubit}$ is the Pauli-$x$ matrix for spin-$1/2$ particle,  $\sigma_x^{(i),env}$ is the Pauli-$x$ matrix for spin-$1$ particle, and $g$ is the coupling constant between the system-qubit and the environment-qutrit. The quantities $\omega_i, \omega_{ij}, \Delta$ and $g_k$ are all in units of time$^{-1}$, $\forall~ i,j$.

{$H_{\mathrm{qubit-env}}$ has multiple degenerate ground states such as $\ket{+++}_q\otimes\ket{+++}_{env}$ and $\ket{-+-}_q\otimes\ket{-+-}_{env}$ where  $\{\ket{+}_q,\ket{-}_q\}$ and $\{\ket{+}_{env},\ket{-}_{env}\}$ pairs are the +1 and -1 eigenstates of $\sigma^{qubit}_x$ and $\sigma^{env}_x$.}
{For the environmental qutrit, the operator $\sigma_x^{\text{env}}$ couples only adjacent energy levels, namely $\ket{2}_{env} \leftrightarrow \ket{1}_{env}$ and $\ket{1}_{env} \leftrightarrow \ket{0}_{env}$. Since there is no matrix element connecting $\ket{2}_{env}$ and $\ket{0}_{env}$, direct transitions between these two states are forbidden. Consequently, the transitions available during a collision depend on the initial state of the qutrit: If the qutrit is initially in the state $\ket{2}_{env}$, only the transition $\ket{2}_{env} \leftrightarrow \ket{1}_{env}$ can occur. If the qutrit is initially in the state $\ket{1}_{env}$, both transitions $\ket{1}_{env} \leftrightarrow \ket{2}_{env}$ and $\ket{1}_{env} \leftrightarrow \ket{0}_{env}$ are allowed. If the qutrit is initially in the state $\ket{0}_{env}$, only the transition $\ket{0}_{env} \leftrightarrow \ket{1}_{env}$ can occur. Thus, the particular qutrit transition is determined solely by the initial qutrit state and the selection rules imposed by $\sigma_x^{\text{env}}$.The qubit operator
$\sigma_x^{\mathrm{qubit}}
= \ketbra{1}{0}_{q}
+ \ketbra{0}{1}_{q}$
enforces that the qubit must flip for the corresponding interaction matrix
element to be nonzero. Here $\ket{1}_q$ and $\ket{0}_q$ are the computational basis states of $2 \times 2$ dimensional Hilbert space.  However, it does not determine which specific qutrit
transition occurs; it only requires that one of the transitions allowed by $\sigma_x^{\mathrm{env}}$ takes place simultaneously.}

The environment is modeled as an infinite sequence of three-level systems or qutrits, each of which sequentially interacts with a single qubit comprising the working substance. Each qutrit interacts with the qubit for a fixed duration \(\delta t\), after which it is replaced by a new, identical qutrit. As a result, the environment effectively resets after each interaction interval \(\delta t\). 
{If we consider the initial state of the system to be $\rho_{in}$ and that of the environmental qutrit to be $\rho_{env}$, the system and the environmental qutrit state after interaction through the global unitary for a time $\delta t$ is given by
\begin{equation}
U=\exp(-iH\delta t),\nonumber
\end{equation}
where $H$ is the total Hamiltonian of the system and the environment, given by
\begin{equation}
H=H_{sys}+H_{env}+H_{qubit-env}.\nonumber
\end{equation}
The Hamiltonians $H_{sys}$, $H_{env}$, and $H_{qubit-env}$ are defined previously in Eqs.~\ref{sys_hamiltonian}, \ref{bath-eqn}, and \ref{qubit-env-eqn} respectively. After the first collision, the state of the composite system is
\begin{equation}
\epsilon_1=U\left(\rho_{in}\otimes\rho_{env}\right)U^{\dagger}.\nonumber
\end{equation}
Consequently, the reduced state of the system after the first collision is
$\rho_1=\mathrm{Tr}_{env}[\epsilon_1]$.
Similarly, after the $n^\text{th}$ collision, the state of the composite system is
\begin{equation}
\epsilon_n=U\left(\rho_{n-1}\otimes\rho_{env}\right)U^{\dagger},\nonumber
\end{equation}
and the reduced state of the system becomes
$\rho_n=\mathrm{Tr}_{env}[\epsilon_n]$. \\}
This construction makes it explicit that the reduced dynamics is obtained by applying a unitary evolution to the composite system followed by a partial trace over the environmental degrees of freedom. Consequently, the reduced-state derivative $\dot{\rho}_X(t)$ in the expression of local heat current is derived from this CPTP collision-map description rather than from the Markovian master equation of Eq.~\eqref{C1}. Since the same CPTP map governs each collision, with only the input state changing from $\rho_{n-1}$ to $\rho_n$, the derivation of $\dot{\rho}_X(t)$, and hence the local heat current, remains valid throughout the dynamics.

 Due to this periodic interaction between the system and environmental units, this framework is commonly referred to in the literature as a collisional model. The dynamical nature of the environment—whether Markovian or non-Markovian - is determined by key parameters of the model, such as the coupling strength between the system and the environment and the collision time.
If the interaction time between the system and each environmental qutrit is sufficiently small for a given interaction strength, the environment is able to extract information from the system but lacks the time necessary to return it. This results in Markovian dynamics, characterized by a lack of memory effects. Conversely, when the interaction time with each qutrit is increased, the environment can not only retain information about the system but also feed it back during subsequent interactions. This backflow of information leads to non-Markovian dynamics, wherein memory effects play a significant role. In our collisional model setup, the environment particles interact with the system qubits continuously for a period of time, which attributes to the non-Markovianity in the system. 
On the other hand, if the environmental qutrits interact perturbatively with the system, i.e., the interaction strength is very weak, and are subsequently replaced after each interaction, as prescribed by the collisional model, the environment fails to retain sufficient information to influence the evolution of the system. As a result, the system exhibits Markovian dynamics, characterized by the absence of memory effects. In contrast, when the system–environment interaction strength is increased, the environment becomes more entangled with the system and retains information about its past states. If the structure of subsequent interactions permits, this stored information can flow back into the system, thereby influencing its future dynamics and giving rise to non-Markovian behavior.

\subsection{Relevant heat currents in working-substance thermal transistors}
\label{heat-current-def-sect}
The performance of a QTT depends on how efficiently it can amplify the heat currents in the terminals, $L$ and $R$, while being controlled by the flow of heat current at terminal $M$.
The definition of heat current in the context of quantum transistors 
is given by
the rate of change of energy of the working substance with respect to time, 
which focuses on
the current flowing 
into and out of the individual baths.
This scenario involves the consideration of local Markovian baths, where heat current is defined as
\begin{equation}\nonumber
    \mathcal{J}_{X} = \text{Tr}(H_{sys}\mathcal{L}_X[\rho]), X \in \{L,M,R\},
\end{equation}
where $\rho$ is the density matrix of the working substance, and $\mathcal{L}_X[\rho]$ is the Lindblad operator corresponding to the terminal $X$, that appears in the Gorini-Kossakolski-Sudarshan-Lindblad (GKSL)~\cite{gksl-ref-1,gksl-ref-2,breuer2002theory} master equation for the Markovian dynamics.
We will refer to these transistors as bath thermal quantum transistors (BTTs) since 
in this case, one effectively considers the heat flowing in and out of a single bath into the entire working substance. This heat current is ``global" in a sense that heat flows in and out of the entire system.
We choose to focus on the heat current which flows in and out of each qubit of the working substance,
to and from different parts of the system and environment, 
and refer to such type of transistors as working-substance thermal quantum transistors (WTTs) since the transistor action occurs due to the heat exchange between different parts of the system and environment, and a particular qubit. The local heat currents flowing in and out each $X^{\text{th}}$ terminal
of the working substance is defined by
\begin{equation}
\label{heat-current-def}
    J_{X} = \text{Tr}(\dot{\rho}_{X}H_{X}),
\end{equation}
where $X \in \{L,M,R\}$. The operator, $\dot{\rho}_{X}$, denotes the rate of change of the reduced density matrix of the $X^{\text{th}}$ terminal of the working substance. This type of heat current can be considered ``local'' in a sense that heat flows locally in and out of a particular subsystem of the working substance. The difference between the two approaches is pictorially depicted in Fig~\ref{system}. This definition of heat current is constructed in a generalized setting, and is not derived via any master equation formalism.
{The energy of a quantum system described by the state $\rho$ and Hamiltonian $H$ is given by
$E = \Tr(\rho H)$.
Let the system and the environment be initially prepared in the states $\rho_S(0)$ and $\rho_E$, respectively. During the first collision ($t<\delta t$), the composite state evolves as
$
\rho_{SE}(t)=U(t)\left(\rho_S(0)\otimes\rho_E\right)U^\dagger(t),
\quad
U(t)=e^{-iHt},
$
where $H$ is the total Hamiltonian. The reduced state of the system is obtained by tracing out the environment,
$
\rho_S(t)=\mathrm{Tr}_E[\rho_{SE}(t)],
$
and the reduced state of the $X^{\text{th}}$ subsystem is
$
\rho_X(t)=\mathrm{Tr}_{\bar{X}}[\rho_S(t)].
$
Therefore,
$
\dot{\rho}_X(t)=\mathrm{Tr}_{\bar{X}}[\dot{\rho}_S(t)].
$
The local energy of the $X^{\text{th}}$ subsystem is then defined as
$
E_X(t)=\mathrm{Tr}[\rho_X(t)H_X].
$
The heat current is defined as the rate of change of energy. Therefore, the local heat current $J_X$ associated with the subsystem $\rho_X$ is given by}
{\begin{eqnarray}\nonumber
J_X &=& \frac{\mathrm{d}}{\mathrm{dt}} E_X \
= \frac{\mathrm{d}}{\mathrm{dt}} \Tr(\rho_X H_X) \\\nonumber
&=& \Tr(\dot{\rho}_X H_X + \rho_X \dot{H}_X) \
= \Tr(\dot{\rho}_X H_X),
\end{eqnarray}
since $\dot{H}_X = 0$.
We arrive at the last equality because the local Hamiltonian $H_X$ is time-independent.
The above derivation has been presented for the first collision interval, but it applies identically to every subsequent collision. After the $n^{\text{th}}$ collision, the system state evolves as
$\rho_n=\mathrm{Tr}_{\mathrm{env}}\!\left[U\left(\rho_{n-1}\otimes\rho_{\mathrm{env}}\right)U^\dagger\right].
$
Thus, each collision is governed by the same CPTP map, with only the input state updating from $\rho_{n-1}$ to $\rho_n$. Consequently, the derivation of $\dot{\rho}_X(t)$, and hence the local heat current,
$
J_X(t)=\mathrm{Tr}\!\left[\dot{\rho}_X(t)H_X\right],
$
remains valid throughout the collision dynamics.

This establishes Eq.~\eqref{heat-current-def}. It is to note that this expression follows directly from the definition of the local energy and is therefore independent of the specific environmental model employed. In particular, its validity does not rely on the collision-model framework, any master-equation description, or any Markovian approximation. The quantity $J_X$ simply quantifies the instantaneous rate of change of the local energy of subsystem $X$.}

In the fabrication of quantum devices, the material properties of the qubits, such as melting point, are typically well-characterized in advance and require minimal information about the surrounding environment. From this standpoint, analyzing heat currents from the perspective of the qubits themselves, rather than from the conventional bath-centric viewpoint, enables a direct correlation between these currents and the intrinsic material properties that are readily accessible in experiments. Such a system-focused definition of heat current has the potential to guide the design and optimization of qubit-based quantum devices whose performance is sensitive to the behavior of these currents.

In our work, the dynamics of the working substance is modeled as general CPTP dynamics, realized through global unitary evolution of the combined system and environment, followed by tracing out the environment. This framework goes beyond the Markovian assumption. Therefore our approach does not rely on master equations, neither global nor local  and as the system does not relax to a steady state. Our numerical analysis further confirms that the system exhibits continuous time-dependent evolution without approaching a steady state. Consequently, since no steady state is attained, the currents do not sum to zero as in their classical counterparts. Within this setting, the definition of heat current that we introduce is novel, as it captures the flow of heat through the system even in the absence of a steady state.

To find the amplification factor, 
we calculate the first order partial derivatives numerically using the five-point formula. According to the five-point formula, the first order derivative of a function $f(x)$ with respect to $x$ at point $x=x_0$ is given by
\begin{multline}
    f'(x_0) \\
     =\frac{f(x_0-2h) - 8f(x_0-h) + 8f(x_0 + h) - f(x_0 +2h)}{12h},
     \label{presc}
\end{multline}
where $h$ is a small real number that is taken to be equal to $0.05$ in our calculations. The five-point formula introduces an error of $O(h^4)$ which is equal to $6.25 \times 10^{-6}$ in our case.

\begin{figure*}[htbp]
    \subfigure[]{\includegraphics[width=0.33\textwidth]{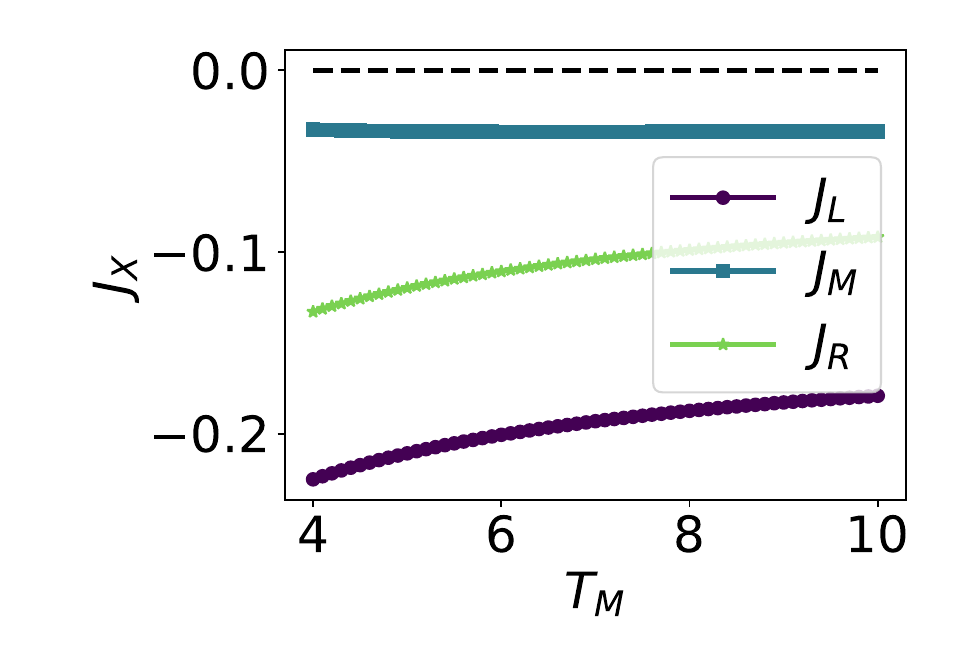}} 
    \label{a}
    \subfigure[]{\includegraphics[width=0.33\textwidth]{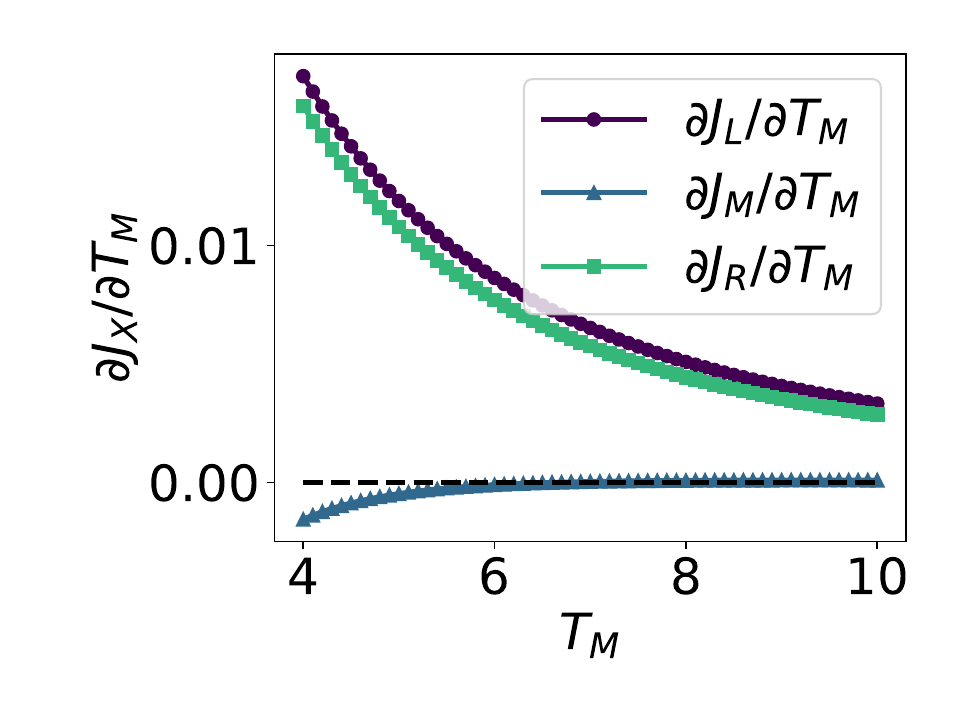}} 
    \label{b}
    \subfigure[]{\includegraphics[width=0.33\textwidth]{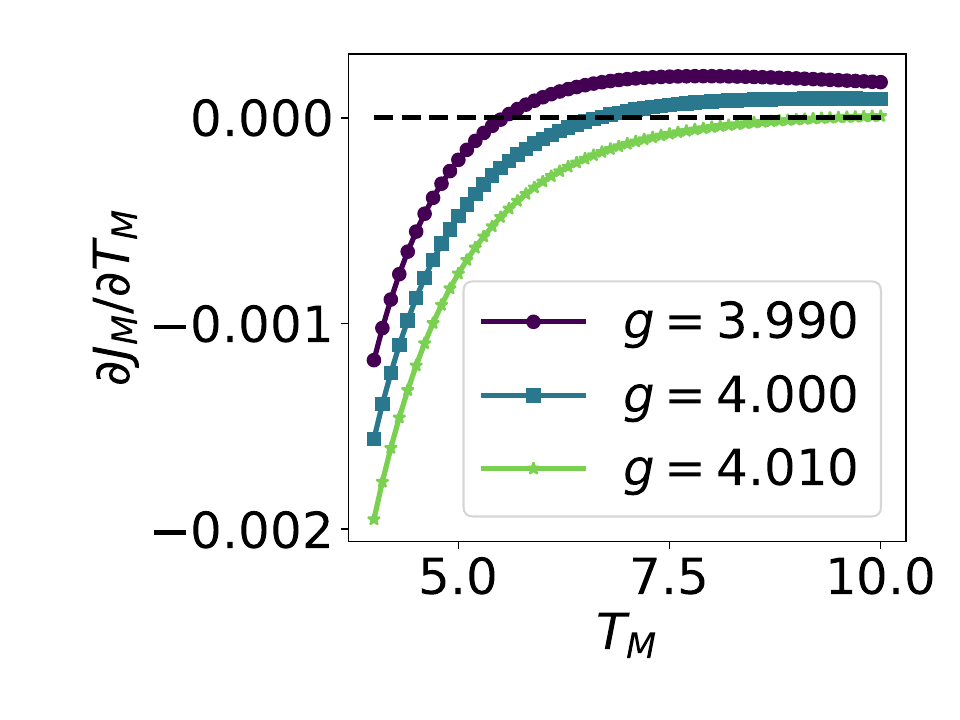}} 
    \label{c}
    \caption{(a) Variation of heat currents at time = 1~$\tilde{t}$  (b) Variation of $\partial J_X / \partial T_M$ with $T_M$, for $X\in\{L,M,R \}$, is depicted. Values of $\partial J_M / \partial T_M$ are much smaller than the other two, and becomes equal to zero at a particular value of $T_M$ which is equal to $6.65$ in units of $\Tilde{T}$ when $g = 4$. (c) Variation of $\partial{J_M}/\partial{T_M}$ with change in the coupling constant between the qubit and the environment is demonstrated. We see that, for every value of $g$, the quantity $\partial{J_M}/\partial{T_M}$ becomes zero at some value of $T_M = T_M^{critical}$, which makes the amplification factor undefined. This value of $T_M^{critical}$ increases as we increases the value of $g$. The black dotted line is for $\partial{J_M}/\partial{T_M} = 0.$ The horizontal axis has dimension $\tilde{T}$ while the vertical axis when multiplied by $\hbar/\Tilde{T}$ will have the dimension of ratio of rate of change of energy with temperature.}
    \label{currents-and-dervs}
\end{figure*}

\section{WTT in presence of linear environment}
\label{linear-sec}
In this section, we work with linear environment and consider the parameter space which allows the WTT to produce amplification. 
The temperature of the middle environment is taken to be in between that of the left and right environments. We consider the initial state of the system comprising the three qubits to be 
$\ketbra{000}{000}$, where $\ket{0}$ denotes the eigenstate of Pauli-$z$ with eigenvalue $+1$. The initial state of the environment $X$ is considered to be the thermal state, 
given by
\begin{equation}
    \rho_{env}^X = \frac{e^{\hbar \Delta \beta_X  \sigma_z^{(X),env}}}{\text{Tr}(e^{\hbar \Delta \beta_X \sigma_z^{(X),env}})},
\end{equation}
where $\beta_X = 1/(k_B T_X)$ is the inverse temperature of the environment $X$ in units of energy$^{-1}$, for $X \in \{L,M,R\}$.

We have defined time as $t~\Tilde{t}$ where $t$ is a dimensionless quantity whose magnitude denotes the magnitude of time, and $\Tilde{t}$ has the dimension of time with magnitude equal to $1$. Similarly, temperature is defined as $T~\Tilde{T}$ where $T$ is dimensionless quantity denoting the magnitude of temperature, and $\Tilde{T}$ has the dimension of temperature with magnitude equal to $1$. $\Tilde{t}$ and $\Tilde{T}$ are related by the relation
\begin{equation}\nonumber
    \Tilde{T} = \frac{\hbar}{k_B~\Tilde{t}},
\end{equation}
where $k_B$ denotes the Boltzmann constant and $\hbar=h/2\pi$, with $h$ being the Planck's constant.

The initial state of the system is considered to be
$\ketbra{000}{000}$, which is the ground state of the system Hamiltonian, $H_{sys}$. 
As we begin with the ground state of the system, it is expected that, during initial time evolution, heat 
will flow from the environment to the system. 
The sign convention adhered to in our work is that 
a positive heat current implies a flow of current from the system to the environment.
This is what we numerically observe as well. The currents are initially 
negative at a small timescale
$\sim 1 \tilde{t}$.
%
{Since the dynamics generated by the collision model does not reach a stationary state, the conventional steady-state energy conservation condition is not expected to hold. Moreover, the total energy balance acquires an additional contribution arising from the interaction energy between the qubits. Therefore, the local currents considered here should not be interpreted as satisfying the usual steady-state conservation relation, even though the overall energy conservation, including the contribution from the interaction term, remains well defined throughout the transient dynamics. A detailed discussion is provided in Appendix~\ref{app_cc}.}


The parameter range in which we perform the analyses is  $T_L = 4 \Tilde{T}, T_R= 10 \Tilde{T}$, $\Delta = 3\Tilde{t}^{-1}, g = 4\Tilde{t}^{-1}$, the time of interaction of the system-qubit with the environment-qutrit is $\delta t = 0.5\Tilde{t}$, and the time at which amplification is measured is $t= 1\Tilde{t}$, unless specified. 
In Fig.~\ref{currents-and-dervs}, we demonstrate the behavior the derivatives of heat currents with respect to temperature, while operating within this particular parameter regime. Fig.~\ref{currents-and-dervs}-(a) shows that, $\partial J_M/ \partial T_M$, the temperature  derivative of the local current of middle qubit, is smaller than that of those corresponding to other qubits, within a temperature range, $4\Tilde{T}$ to $10\Tilde{T}$. 

This enables us to obtain amplification within a sufficiently large temperature range for the middle environment. 
It is to be noted that we have chosen our parameter regime where the coupling strength between the environment and the working substance is comparable to the coupling strength between the sub-parts of the working substance.
 We see in panel Fig.~\ref{currents-and-dervs}-(a)  that since $\partial J_M / \partial T_M$ is much smaller than the other two current derivatives in this parameter space, we get a good amplification. We also note that there exists a parameter region where $\partial J_M / \partial T_M  = 0$. Here the amplification {diverges}. 
In panel (b) of Fig.~\ref{currents-and-dervs}, we have shown the variation of $\partial J_M / \partial T_M$ with $T_M$ for different values of system-environment coupling $g$.



  


In the succeeding subsection, we study the amplification of the heat currents flowing into the the left and right qubits in presence of local environments connected to each of the qubits separately. In subsection~\ref{sub2}, we analyze the thermal transistor effects with less than three local environments connected to three system qubits.

\subsection{Amplification with three local environments}
\label{sub1}
In this section, we consider the scenario where the system comprises three qubits, and each qubit is connected with a local environment that pertains to the collisional model composed of periodically interacting qutrits.
\subsubsection{Dependence of amplification on the system-environment interaction}

The performance of a WTT depends on how efficiently it can amplify the heat currents at the left and right terminals, while being modulated by the temperature of the middle terminal.
Here we analyze how the amplification at the left and right terminals vary with temperature of the middle qubit by perturbatively varying the interaction strength between the working substance and environment.  
{We further investigate the robustness of the transistor behavior with respect to the temperature of the middle qubit by considering different numbers of interactions and different interaction times.}

\begin{figure*}
    \includegraphics[width=15.8cm, keepaspectratio]{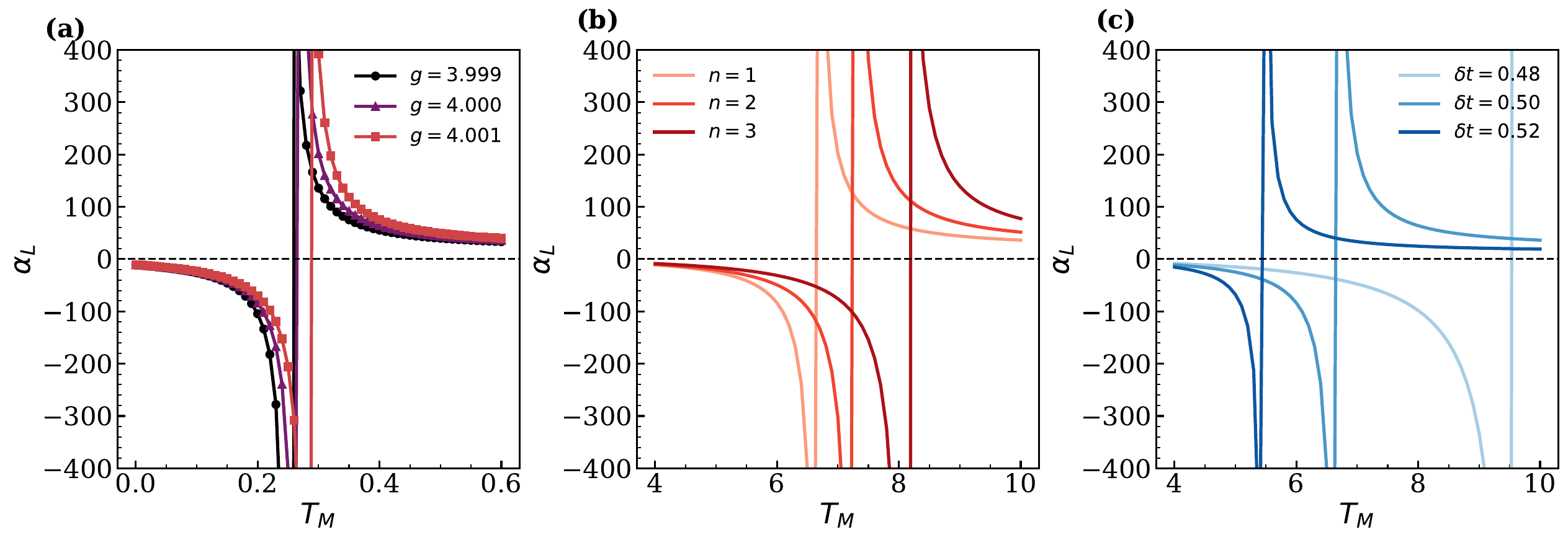}
    \caption{Panel (a) shows the variation of the dynamical amplification factor, $\alpha_L$, along the vertical axis, versus the temperature of the middle bath, $T_M$, along the horizontal axis, for different values of system-bath interaction strength, $g$. 
    We observe that there is a discontinuity in the value of $\alpha_L$ with respect to $T_M$, and an increase in the interaction strength increases and decreases the amplification to the right and left of the discontinuity respectively.
    The black dotted line signifies zero amplification. The quantity, $\alpha_L$, is dimensionless, while $T_M$ is in units of $\tilde{T}$. {Panel~(b) presents the same analysis for $g=4$ and $\delta t=0.5$, considering different numbers of interactions, $n=2$, $3$, and $4$. Panel~(c) shows the variation of $\alpha_L$ with $T_M$ for $g=4$ and $n=2$, for three different interaction times, $\delta t=0.48$, $0.50$, and $0.52$.}}
    \label{alpha-L-var-gk}
\end{figure*}

Panel (a) of Fig.~\ref{alpha-L-var-gk} illustrates the dynamical amplification factor at the left terminal, $\alpha_L$, as a function of the middle bath's temperature, $T_M$, for different perturbative variations in the system-environment interaction strength, $g$.
 In this regard, we find that for a given interaction strength, the magnitude of amplification first increases until a certain value of $T_M$. Let us refer to this value as $T_M^{critical}$. At this particular temperature, $T_M^{critical}$, the dynamical amplification factor $\alpha_L$, diverges. 
The divergence occurs as a result of $\partial{J_M}/\partial{T_M}$ approaching zero, indicating a point of vanishing sensitivity of the current to changes in the temperature of the middle bath.
This feature is depicted in Fig.~\ref{currents-and-dervs}-(b), where the derivative of the current with respect to $T_M$ is shown to diminish around $T_M^{critical}$. It is also observed that, at this point of discontinuity, the dynamical amplification factor, $\alpha_L$, switches its sign from negative to positive. 
Towards the right of $T_M^{critical}$, i.e. where $T_M > T_M^{critical}$, the magnitude of amplification decreases and saturates to a value which is greater than one. 
This illustrates that transistor effect is also present for relatively higher values of $T_M$.
The value of $T_M^{critical}$ increases with the increase of system environment interaction strength. This feature can also be understood from Fig.~\ref{currents-and-dervs}-(b), where the value of $T_M$ for which the $\partial{J_M}/\partial{T_M}$ curve cuts the abscissa increases with increasing interaction strength.
Furthermore, we show that, in the region $T_M < T_M^{critical}$, a lower interaction strength results in a higher magnitude of amplification being obtained for a particular value of $T_M$, whereas this trend is reversed in the $T_M > T_M^{critical}$ region, where a higher interaction strength results in higher amplification. 
%
%
The behavior of $\alpha_R$ with respect to $T_M$ is not shown because its behavior is identical to that of $\alpha_L$, and this occurs due to the symmetry between the left and right terminals of the setup.

%

In Fig.~\ref{currents-and-dervs} (a), we show the heat current flowing through the $X^{\text{th}}$ qubit, $J_X$, is plotted along the vertical axis, as a function of the middle-bath temperature, $T_M$ along the horizontal axis. Here we see that compared to the variation in $J_L$ and $J_R$, the quantity $J_M$ varies very slowly with respect to $T_M$. This feature is also evident from Fig.~2(b), where see that the quantity, ${\frac{\partial J_M}{\partial T_M}}$, is much smaller than ${\frac{\partial J_L}{\partial T_M}}$ and ${\frac{\partial J_R}{\partial T_M}}$. Therefore, the denominator of the dynamical amplification factor becomes significantly higher than its numerator, resulting in an appreciable amplification in the parameter regime in which we work. Let us now qualitatively interpret the behavior of the amplification factor, $\alpha_L$, as given in Fig.~3, by looking at the heat-current derivatives given in Figs.~2(b) and 2(c). From Fig.~2(c), we see that in the temperature regime, $T_M<T^{critical}_M $, the quantity, ${\frac{\partial J_M}{\partial T_M}}<0$. Then it gradually increases to $T_M=T^{critical}_M $, where ${\frac{\partial J_M}{\partial T_M}}=0$ and in the regime $T_M>T^{critical}_M$, the quantity, ${\frac{\partial J_M}{\partial T_M}}>0$. Further, from Fig.~2(b) we observe that the quantity, $\frac{\partial J_L}{\partial T_M}>0$, $\forall$ $T_M$. 
Therefore, for small values of $T_M$, the value of ${\frac{\partial J_L}{\partial T_M}}$ is positive and ${\frac{\partial J_M}{\partial T_M}}$ is negative, which causes negative amplification.  
At $T_M=T^{critical}_M$, the quantity, ${\frac{\partial J_M}{\partial T_M}}=0$, which causes the amplification factor, $\alpha_L$, to diverge at the critical value of $T_M=T^{critical}_M$. Finally, for values of temperatures of the middle bath, $T_M>T^{critical}_M$, both ${\frac{\partial J_M}{\partial T_M}}$ and ${\frac{\partial J_L}{\partial T_M}}$ are positive, which results in positive amplification.
{It is to note that the heat current $J_X(t)$ and amplification are computed continuously during the evolution rather than only after the completion of a collision event.}

{The physical origin of the large amplification can be understood directly from the expression of the dynamical amplification factor. As shown in Fig.~\ref{currents-and-dervs}, the variation of the middle-terminal heat current with respect to the control temperature, namely $\partial J_M/\partial T_M$, becomes significantly smaller than the corresponding variations of the left and right heat currents, $\partial J_X/\partial T_M$. Consequently, the denominator of Eq.~(\ref{amp-factor-eqn}) remains much smaller than the numerator, resulting in a large amplification factor at a specific value of $T_M$. Importantly, the transistor effect in our model is not restricted to a singe operating point. The physically relevant observation is that, in the vicinity of the vey high amplification point, the amplification factor remains finite but very large. As shown in panel (a), (b) and (c) of Fig.~\ref{amp-factor-eqn}, $\alpha_L$ reaches values of the order of $10^2$ over an extended parameter range and not solely at the point where the denominator vanishes. 
}

{We also perform an analysis by varying both the number of interaction steps and the collision duration $\delta t$.  
Specifically, in Fig.~\ref{alpha-L-var-gk}-(b), we plot the amplification factor $\alpha_L$ as a function of $T_M$ for different numbers of interaction steps, i.e. $n=2,3,$ and $4$. In all cases, the system exhibits substantial amplification, and the qualitative behavior of the amplification curve remains unchanged. In this study $\delta t$ is fixed at 0.5. Furthermore, in Fig.~\ref{alpha-L-var-gk}(c), we present the corresponding results for different collision durations, i.e. $\delta t=0.48$, $0.50$, and $0.52$. We find that the transistor effect persists for all these values of $\delta t$, with only minor quantitative variations. Here the number of interactions is fixed at $n=2$. These results demonstrate that  the occurrence of significant thermal amplification and transistor-like behavior is robust with respect to variations in both the number of collisions and the collision duration.} {In Appendix~\ref{app_init_states}, we further demonstrate that the transistor effect is not exclusive to the initial state $\ket{0}^{\otimes 3}$, but also arises for other choices of initial states.}

{In our calculations, the derivatives appearing in Eq.~\eqref{presc} are evaluated using a five-point finite-difference formula, whose truncation error scales as $\mathcal{O}(h^4)$. Throughout the {work}, we use a step size of $h=0.01$, resulting in an estimated numerical error of the order of $10^{-8}$. We have verified that, within the entire temperature range considered in {work}, the minimum value of $\left|\partial J_M/\partial T_M\right|$ remains of the order of $10^{-4}$. Therefore, the numerical error associated with the derivative evaluation is approximately four orders of magnitude smaller than the smallest denominator encountered in the amplification factor. This indicates that the large amplification observed in our results is not a numerical artifact arising from the finite-difference scheme, but rather reflects the underlying physical behavior of the system.}


\subsubsection{Variation of amplification with time}
\begin{figure}
    \centering
    \includegraphics[width=\columnwidth, keepaspectratio]{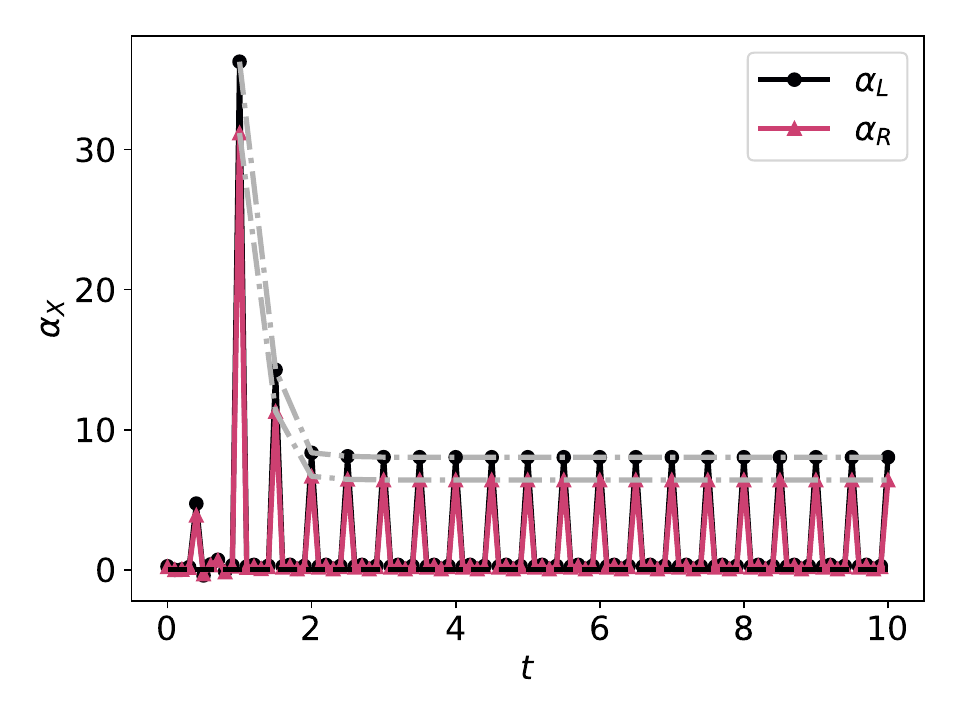}
    \caption{Variation of the dynamical amplification factor, $\alpha_X$, with time, $t$, at $T_M = 10$. Here $X\in L,R$ where $L$ and $R$ denote left and right qubits respectively. We find that the variation becomes periodic after certain number of interactions have happened between the working substance and the environment. The dotted black line signifies amplification equal to zero. The vertical axis is dimensionless and the horizontal axis is in units of $\tilde{t}$.}
    \label{amp-w-time}
\end{figure}

In this section, we analyze
the variation of dynamical amplification factor of a WTT as a function of time for
a given value of system-environment coupling and a fixed temperature of the environment connected to the middle qubit.
%
%
We find that the variation of the dynamical amplification factor with time becomes periodic after the occurrence of a certain number of interactions between the working substance and the environment. After a finite number of interactions, 
we 
find that the amplification shoots up and falls down periodically, as shown in  Fig.~\ref{amp-w-time}.

{The amplification factor exhibits a sequence of oscillatory peaks as a function of time. The envelope obtained by joining the successive maxima decreases monotonically, whereas the instantaneous amplification displays pronounced non-monotonic oscillations. Such oscillatory behavior alone, however, does not constitute evidence of non-Markovianity, since periodic collision protocols can also induce similar features.}

{To distinguish genuine memory effects from simple periodicity, we characterize the dynamics using the Breuer-Laine-Piilo (BLP) measure of non-Markovianity~\cite{BLP}. The nonzero values of the BLP measure obtained for the present dynamics demonstrate the existence of information backflow from the environment to the system and thereby provide an independent operational witness of non-Markovian behavior. Furthermore, the increase of the BLP measure with evolution time indicates a progressive strengthening of memory effects in the system dynamics.}

{The BLP measure associated with a quantum CPTP map $\Phi$ is defined as
\begin{equation}
\mathcal{N}(\Phi)
=
\max_{\rho_{1,2}(0)}
\int_{\sigma>0} dt \,
\sigma\!\left(t,\rho_{1,2}(0)\right),
\end{equation}}
{where the integration is restricted to the time intervals for which
$\sigma\!\left(t,\rho_{1,2}(0)\right)>0$, and the maximization is performed over all possible pairs of initial states $\rho_1(0)$ and $\rho_2(0)$. The quantity, $\sigma\!\left(t,\rho_{1,2}(0)\right)$, is defined as}
{\begin{equation}
\sigma\!\left(t,\rho_{1,2}(0)\right)
=
\frac{d}{dt}D\!\left(\rho_1(t),\rho_2(t)\right).
\end{equation}}
{Here, the trace distance, $D(\rho_1,\rho_2)$, is defined as}
{\begin{equation}
D(\rho_1,\rho_2)
=
\frac{1}{2}\,\mathrm{tr}\!\left|\rho_1-\rho_2\right|,
\end{equation}}
{with the usual definition of trace norm, 
$|A|=\sqrt{A^\dagger A}$, for a given operator $A$.} \\

{For the parameter regime corresponding to Fig.~3 and a fixed middle-bath temperature {$T_M=8$,} the computed values of the BLP measure $\mathcal{N}(\Phi)$ are presented in Table~\ref{T_1}. }
\begin{table}[h]
\centering
\begin{tabular}{cc}
\hline
Time ($t$) & BLP value $\mathcal{N}(\Phi)$ \\
\hline
0.1 & 0.513939 \\
0.2 & 0.580511 \\
0.3 & 0.627911 \\
0.4 & 0.658938 \\
0.5 & 0.803348 \\
0.6 & 0.811772 \\
\hline
\end{tabular}
\caption{{Values of the BLP measure of non-Markovianity, $\mathcal{N}(\Phi)$, for different evolution times.}}
\label{T_1}
\end{table}

{The nonzero values of $\mathcal{N}(\Phi)$ provide clear evidence of the non-Markovian nature of the dynamics. Moreover, the monotonic increase of $\mathcal{N}(\Phi)$ with evolution time signifies an increasing degree of information backflow from the environment to the system and, consequently, a progressive strengthening of memory effects. These results quantitatively demonstrate that the dynamics becomes increasingly non-Markovian as time progresses.}

{Therefore, while the non-monotonic behavior of the amplification cannot by itself be regarded as a signature of non-Markovianity, its occurrence in conjunction with a finite BLP measure suggests that the observed amplification dynamics evolves in a genuinely non-Markovian regime.
}


The jumps occur periodically with periodicity $0.5\Tilde{t}$, which is  the time of interaction 
between each qutrit of the environment and the working substance.
As a qutrit pertaining to the environment arrives to interact with a system-qubit, the amplification remains small for a short time initially and then rises abruptly to a higher value. 
The maximum value corresponding to each jump is reached when the qutrit has stopped interacting with a qubit of the WTT. Following this, the amplification drops as the environment-qutrit moves away from the system-qubit. Then another qutrit arrives to interact with the qubit and the entire process is repeated. Here it is interesting to note that, in the time range in which we work, the number of peaks in the amplification versus time curve, as given in Fig.~\ref{amp-w-time}, 
is sufficient large. 
It basically implies a wide working region of the WTT in that particular range of time.

\begin{figure}
    \centering
    \includegraphics[width=\linewidth]{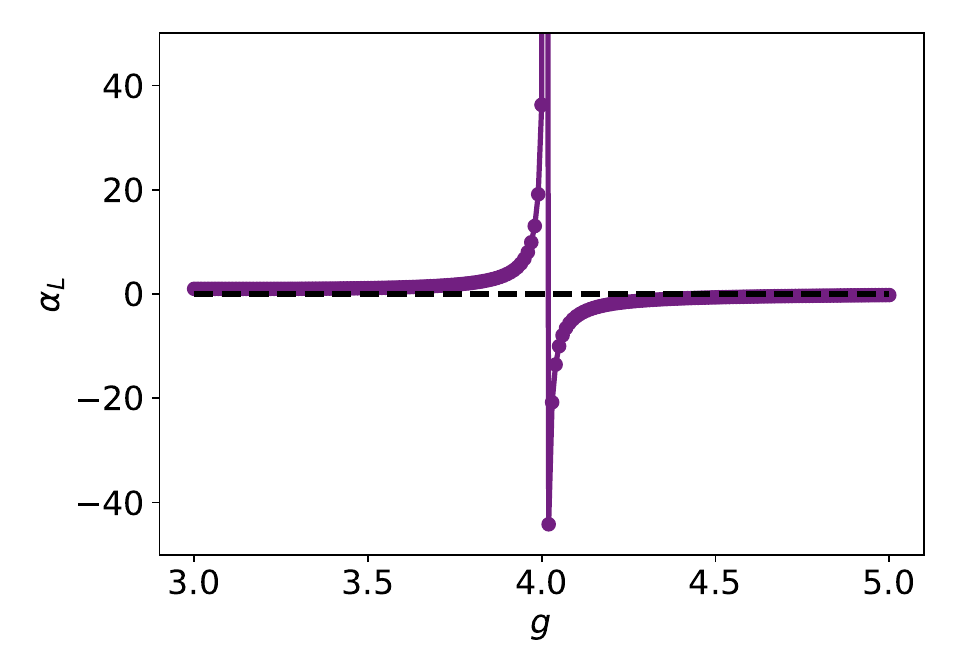}
    \caption{Variation of $\alpha_L$ on the vertical axis as the coupling strength $g$ (on the horizontal axis) between the qubit and the environment is varied at $t = 1\tilde{t}$. We observe that when the coupling strength 
$g<4$, the amplification is small and positive, and when 
$g>4$, it is small and negative. A sharp peak in amplification occurs around 
$g=
4$, indicating a significant enhancement at this point.}
    \label{variation-w-g}
\end{figure}

In order to understand why there is a peak in the amplification versus time plot at $t=1\tilde{t}$, we carried out a perturbative analysis of how the amplification factor behaves around time $t = 1$ within the relevant parameter regime. This revealed some interesting features specific to the Hamiltonian considered in our work. In particular, by varying the coupling strength between each qubit and its respective bath, we observed the following. For coupling strengths slightly below $g=4$, the amplification is small and positive, whereas for coupling strengths slightly above $g=4$, it becomes small and negative. The magnitude of the amplification exhibits a sharp peak at $g=4$, where we observe a pronounced amplification, as illustrated in Fig.~\ref{variation-w-g}. This behavior is an intrinsic feature of the Hamiltonian model in the chosen parameter regime.

\subsection{Amplification with less than three local environments}
\label{sub2}
In this section, we consider the scenario where the system comprises three or less qubits, and the number of environments interacting with the system are less than three.
\subsubsection{WTTs with three qubits and two environments}

To understand the role of the three environments further, we consider the case when we remove one of these environments. We have considered the middle environment, which is the modulating one, to be always present, and we remove either of the left or right environments. The temperatures of the three environments considered for the analysis in this subsection are given by $T_L = 4\Tilde{T}, T_M = 8\Tilde{T}, T_R = 10\Tilde{T}$. 
\begin{figure}
    \centering
    \includegraphics[width=\columnwidth, keepaspectratio]{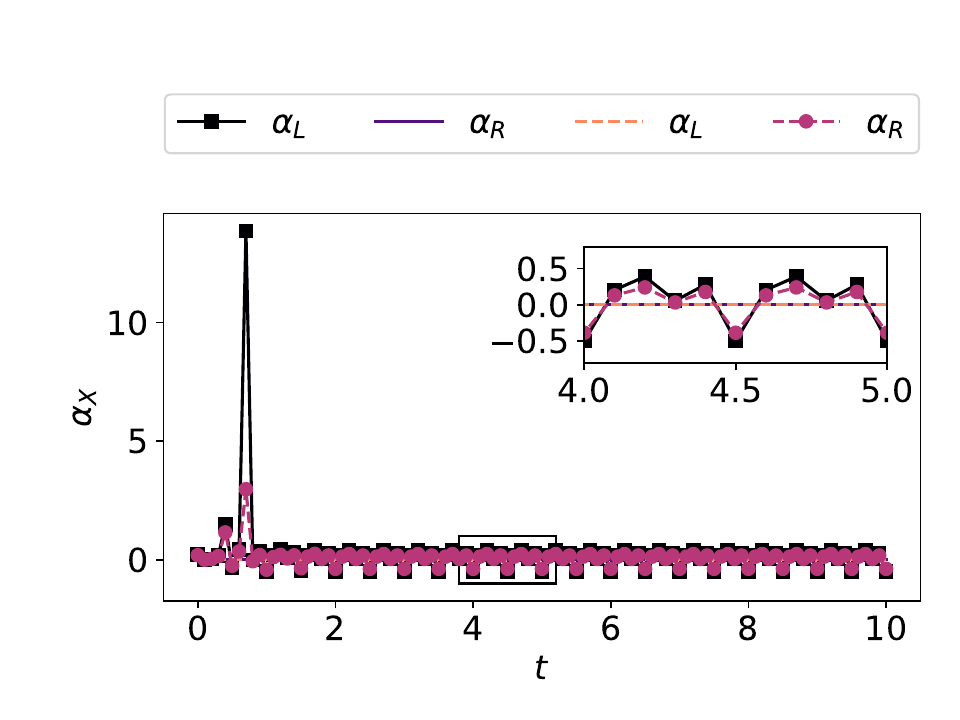}
    \caption{Variation of dynamical amplification factor with time when one of the environments is removed. The solid lines are for the system when the right environment is removed while the dotted lines are for the system when the left environment is removed . The inset shows the variation of amplification in the highlighted portion of the main graph. The vertical axis is dimensionless and the horizontal axis is in units of $\tilde{t}$.}
    \label{amp-one-bath-rem}
\end{figure}

\begin{figure}
    \centering
    \includegraphics[width=\columnwidth, keepaspectratio]{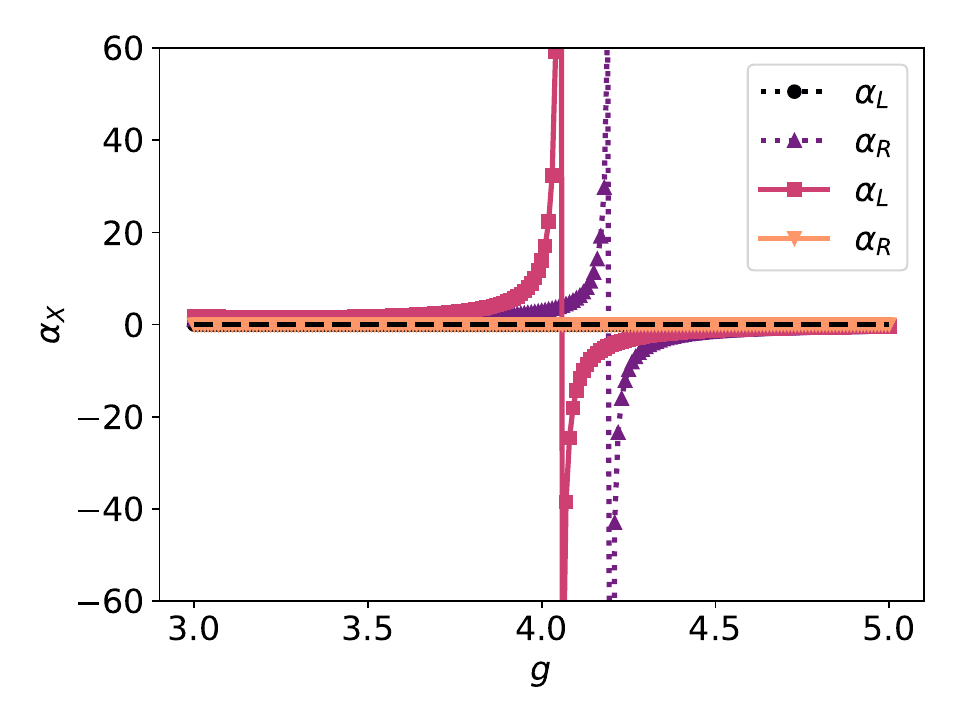}
    \caption{Variation of dynamical amplification factor with coupling strength $g$ when one of the environments is removed. The solid lines are for the system when the right environment is removed while the dotted lines are for the system when the left environment is removed .}
    \label{amp-one-bath-rem-g}
\end{figure}

We first remove the environment connected to the right qubit such that the right qubit only interacts with the two other qubits of the working substance to which it is coupled. We then analyze the variation of the dynamical amplification factor corresponding to the left qubit as a function of time as shown in Fig.~\ref{amp-one-bath-rem}.
The black and purple solid curves depict $\alpha_L$ and $\alpha_R$ in this scenario respectively.
We find that $\alpha_R = \partial J_R / \partial J_M$ becomes zero while $\alpha_L$ takes finite values as time is varied. The value of $\alpha_R$ being equal to zero possibly implies that the change in current happens majorly due to the current flowing into the environments rather than into the other qubits. Since we have removed the right environment, there is no current flowing from the right qubit to the environment, and only the other two qubits are responsible for the current which remains almost zero and remains constant with respect to a change in $T_M$. The global maximum of $\alpha_L$ is attained at 0.7$\Tilde{t}$ seconds with 
the maximum value being $13.85$.


The red dotted curve shows the amplification for the right qubit when the left qubit is disconnected from its environment. On the other hand, the orange dotted curve shows the amplification for left qubit which is disconnected to the environment. The behavior of amplification, $\alpha_L$, versus time is qualitatively similar to the 
case of $\alpha_R$, but with a diminished value in the amplification factor.
We find that $\alpha_R$ reaches maximum
at the end of interaction between the system and the environment. 
This feature
is similar to what we had obtained in the case of three environments attached to the working substance in Fig.~\ref{amp-w-time}, with the period of each jump 
being the same as before,
 i.e. equal to the interaction time, but the maximum value being much lower in this case.
We find that the values of the dynamical amplification factor are greater than 1 and a transistor effect is seen, although the amplification is not very large. The global maximum of $\alpha_R$ in this case is reached at 0.7$\Tilde{t}$ seconds with
the maximum value being $2.97$.


Therefore, when the environment connected to the right qubit is removed, the value of  $\alpha_L$ follows the same qualitative trend as that of $\alpha_R$ when the left bath is removed. But the amplification values are slightly higher in the first case of removing the right bath. The global maxima is reached at the same value of time but is higher and reaches a value of 13.84.  
The disparity between the two cases is because of the two baths having unequal temperatures which affect the amplification in 
each of the cases.


The height of the peak of $\alpha_R$ and $\alpha_L$ in Fig. \ref{amp-one-bath-rem} at time $t=0.7\tilde{t}$ can be explained by analyzing the variation of $\alpha$ with the coupling strength $g$ as shown in Fig. \ref{amp-one-bath-rem-g}. We see that in the case where right environment is removed (corresponding to solid pink and orange lines), the amplification is very small and monotonously increases until it reaches its peak at $g=4.05$ and then suddenly becomes negative. After which the amplitude of the amplification keeps on decreasing. A similar trend is seen when the left environment is removed (corresponding to the dotted black and blue lines). In this case, the peak is reached at a slightly higher $g$ value of $g=4.2$ but the trend remains similar to that of the case when right environment was removed.

We also provide a discussion of the case where the system comprises two qubits, while each qubit is connected to their respective environment, in Appendix~\ref{appA}. 



\subsection{Symmetric versus asymmetric coupling between working-substance qubits} \label{nmsa}

\begin{figure*}[htbp]
    \subfigure[]{\includegraphics[width=0.4\textwidth]{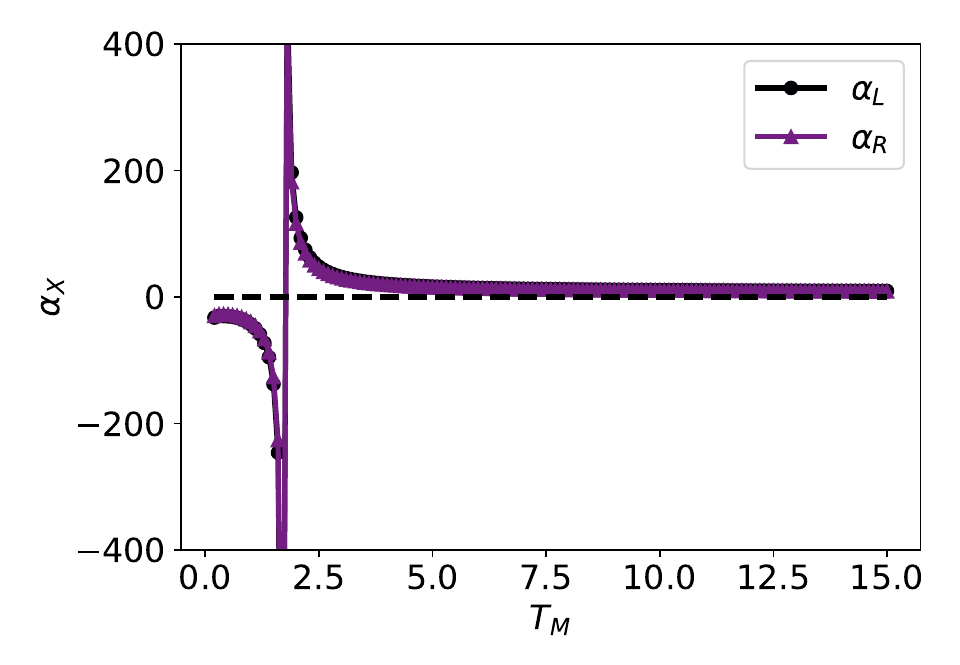}} 
    \label{a}
    \subfigure[]{\includegraphics[width=0.4\textwidth]{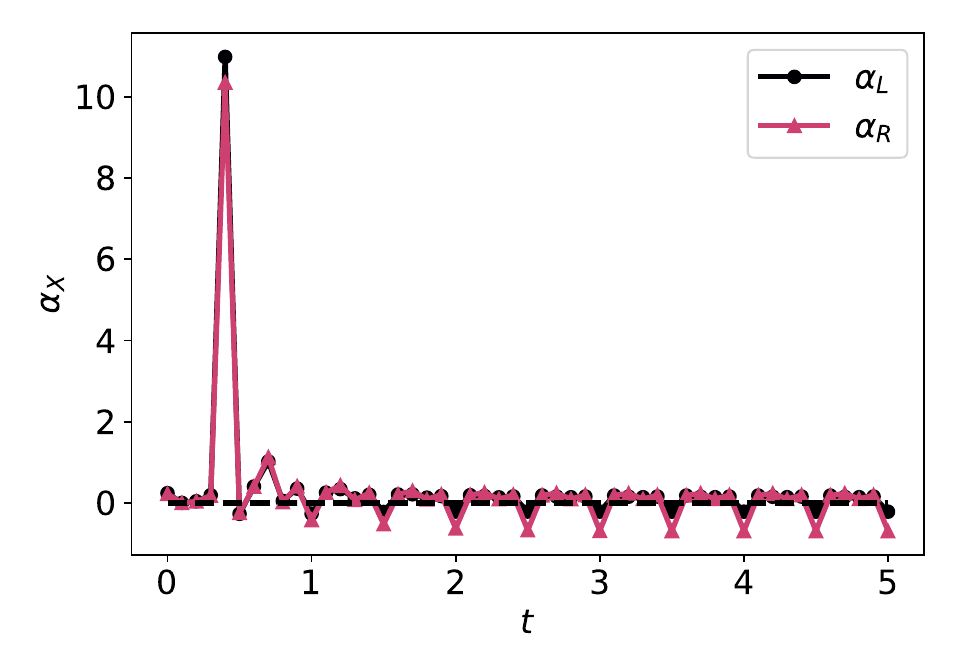}} 
    \label{b}
    \subfigure[]{\includegraphics[width=0.4\textwidth]{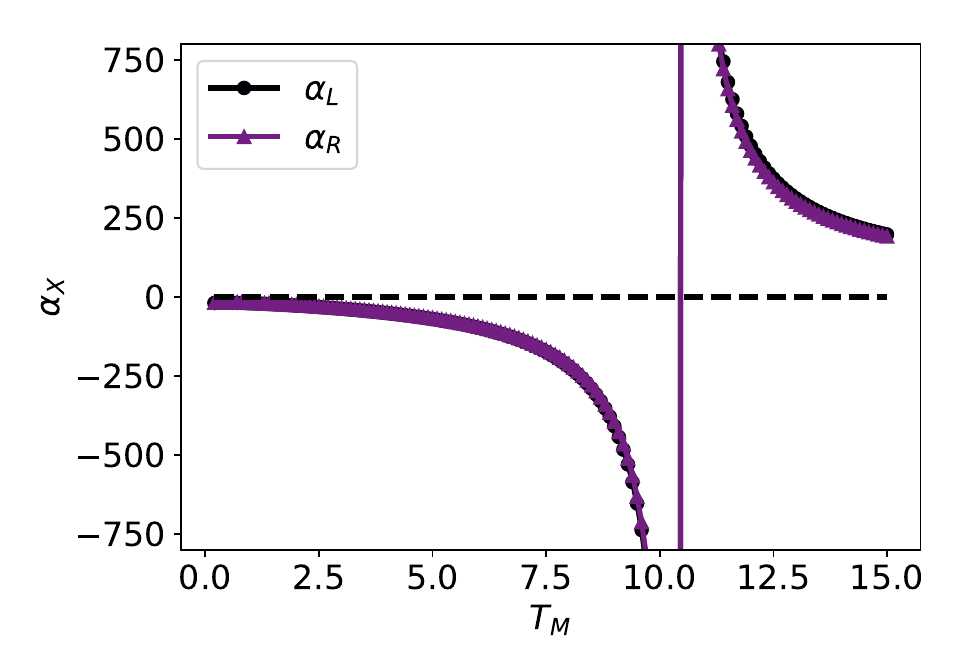}} 
    \label{c}
    \subfigure[]{\includegraphics[width=0.4\textwidth]{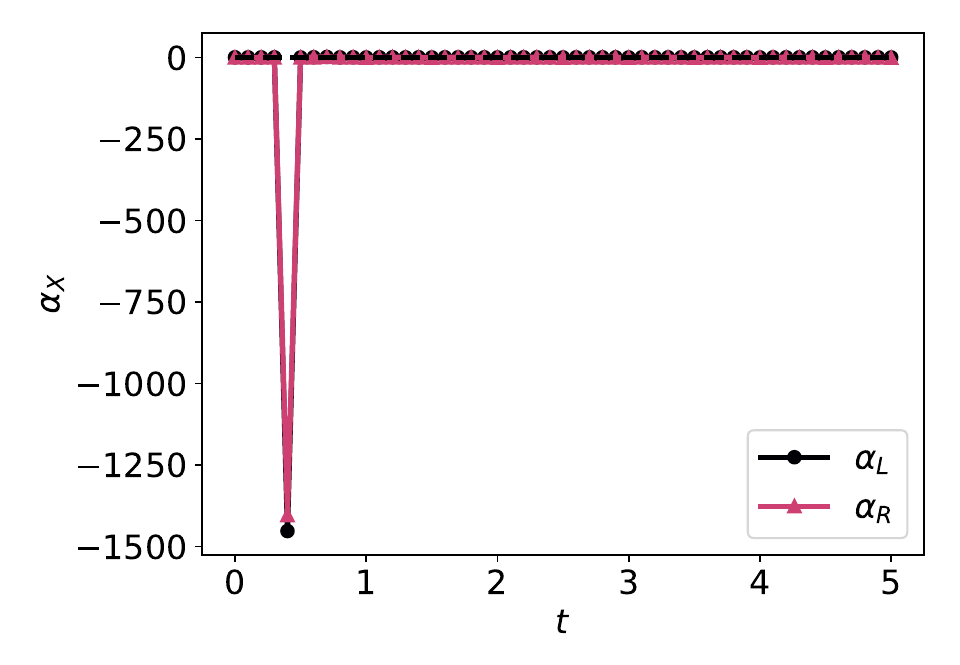}} 
    \label{d}
    \caption{Variation of amplification $\alpha_X$ as temperature of the middle bath $T_M$ and time $t$ change. The top figures (a) and (b) correspond to the symmetric case where the coupling between all the qubits is $\Delta$. The bottom figures correspond to the asymmetric coupling between the pair of qubits. The coupling between left and middle is $\Delta=3$, between middle and right is $\Delta+0.1 = 3.1$ and between left and right is $\Delta-0.1 = 2.9$. (a) and (c) corresponds to time $0.4 \tilde{t}$ while (b) and (d) correspond to $T_M = 10$}
    \label{symm-and-unsymm}
\end{figure*}

In the preceding subsections, we considered the case where there was no coupling between the left and right qubits of the working subsystem, and the couplings between the left–middle and right–middle qubits were equal, i.e
\begin{align}
\label{og-couple}
    \nonumber
    \omega_{ML} &= \omega_{MR} = \Delta \\
    \omega_{LR} &= 0.
\end{align}
Here we consider the following two situations. In the first case, we additionally couple the left and right qubits, and consider the coupling between the left–middle, right–middle and left–right qubits to be equal, i.e.
\begin{align}
\label{symm-couple}
    \omega_{ML} = \omega_{MR} = \omega_{LR} = \Delta.
\end{align}
In the second case, we consider unequal couplings between each pair of qubits, i.e,
\begin{align}
\nonumber
\label{unsymm-couple}
    \omega_{ML} &= \Delta \\
    \nonumber
    \omega_{MR} &= \Delta + \delta \\
    \omega_{LR} &= \Delta - \delta,
\end{align}
where $\delta$ is a parameter considered small compared to $\Delta$. We refer to this case as the asymmetric case. In particular, we have taken $\delta = 0.1$ and $\Delta=3$ as before.

We calculate the dynamical amplification factor for the left and right qubits as functions of the middle bath temperature and the evolution time, for both symmetric and asymmetric coupling scenarios, and present the results in Fig.~\ref{symm-and-unsymm}. Our analysis shows that amplification persists under both types of coupling. The dependence of amplification on the middle bath temperature $T_M$, shown in Figs.~\ref{symm-and-unsymm} (a) and (c) corresponding to symmetric and asymmetric cases, follows a trend similar to that observed previously for the coupling strengths given in Eq.~\ref{og-couple} (refer to Fig.~\ref{alpha-L-var-gk}). Specifically, the amplification is initially negative, with its magnitude increasing up to a critical temperature $T_{M,A(S)}^{\text{critical}}$, beyond which it becomes positive and subsequently decreases. The value of $T_M^{\text{critical}}$ differs between the symmetric and asymmetric cases, with the critical temperature for symmetric case being $T_{M,S}^{\text{critical}} = 1.75$ and for the asymmetric case being $T_{M,A}^{\text{critical}} = 10.45$ 

The time dependence of the amplification, shown in Figs.~\ref{symm-and-unsymm} (b) and (d), also exhibits distinct features. After $t = \tilde{t}$, the amplification becomes periodic with the same period as the interaction time, consistent with the behavior reported in Fig.~\ref{amp-w-time}. In both symmetric and asymmetric cases, the amplification reaches its peak at $t = 0.4 \tilde{t}$.

\begin{figure*}[htbp]
    \subfigure[]{\includegraphics[width=0.33\textwidth]{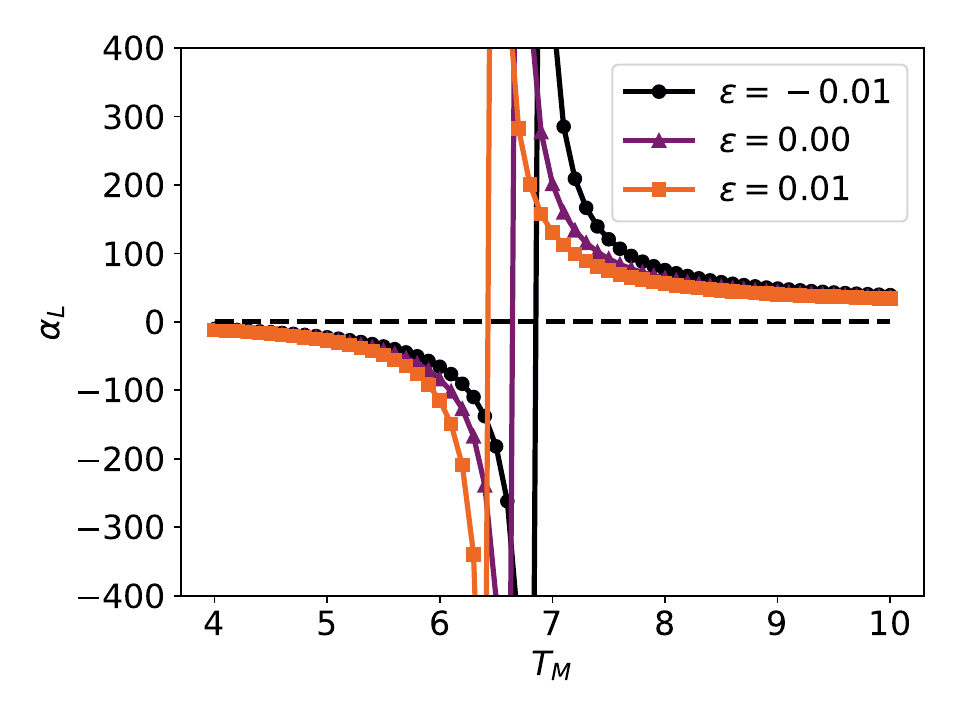}} 
    \label{a}
    \subfigure[]{\includegraphics[width=0.33\textwidth]{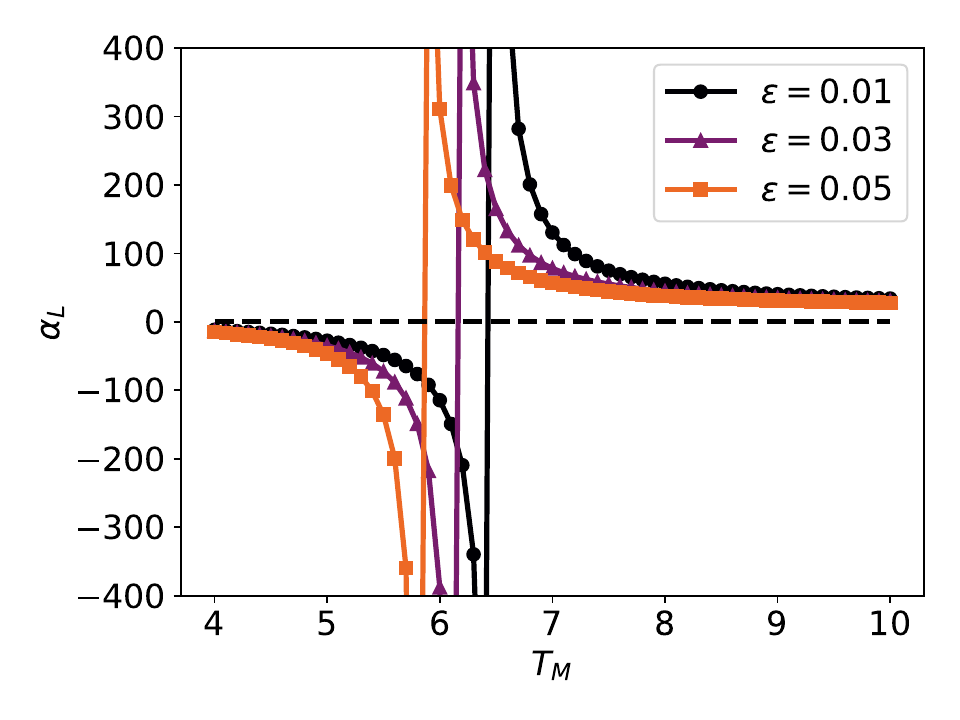}} 
    \label{b}
    \subfigure[]{\includegraphics[width=0.33\textwidth]{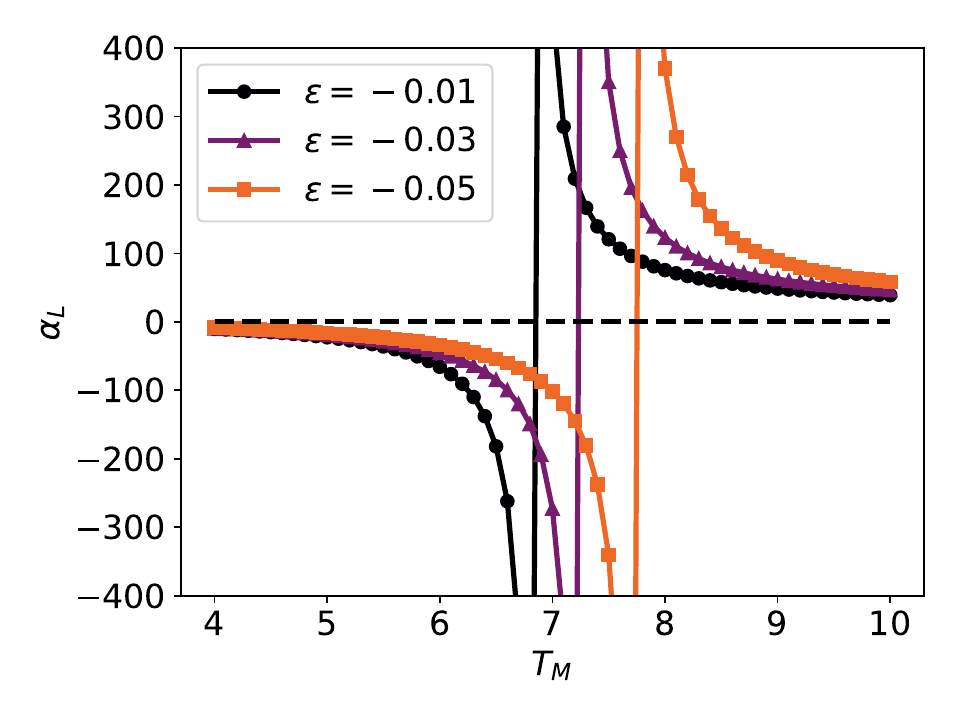}} 
    \label{c}
    \caption{Variation of the dynamical amplification factor, $\alpha_L$, along the vertical axis, versus the temperature of the middle bath, $T_M$, along the horizontal axis, for different values of 
    non-linearity, given by $\epsilon$.
    In panel (a), variation of amplification ($\alpha_L$) with temperature of the middle bath, in the linear case along  with the introduction of Kerr type and transmon non-linearities is given. The curves are obtained by taking time $t=1 \Tilde{t}$. 
    The variation of amplification ($\alpha_L$) with temperature of the middle bath as we increase the non-linearity in the Kerr and transmon type environments at time $t=1 \Tilde{t}$ are plotted in panels (b) and (c) respectively. 
    The quantity, $\alpha_L$, is dimensionless, while $T_M$ is in units of $\tilde{T}$.}
    \label{non-linearity-graph}
\end{figure*}

\begin{figure}
    \centering
    
    \includegraphics[width=\columnwidth, keepaspectratio]{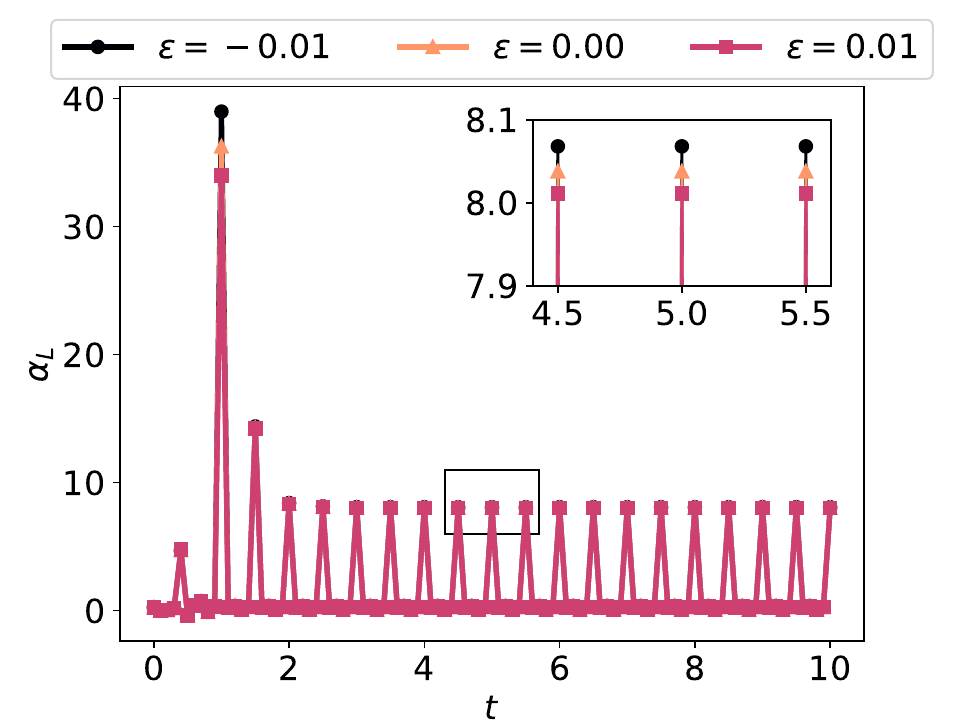}
     \caption{Variation of the dynamical amplification factor, $\alpha_L$, with time, $t$. 
     The inset shows the minute differences between the amplification in the case of transmon, Kerr, and linear environments in the range of time highlighted in the main graph. The vertical axis is dimensionless and the horizontal axis is in units of $\tilde{t}$. The given graph is at $T_M = 10$ in units of $\Tilde{T}$.}
    \label{amp-time-non-linear}
\end{figure}

Furthermore, instead of considering the environment to be a collision model composed of qutrits, we consider the collision model to consist of a qubit system and again show sufficient amplification corresponding to both right and left qubits in Appendix~\ref{qubit_env}.

\begin{figure}
    \centering
    \includegraphics[width=0.80\columnwidth]{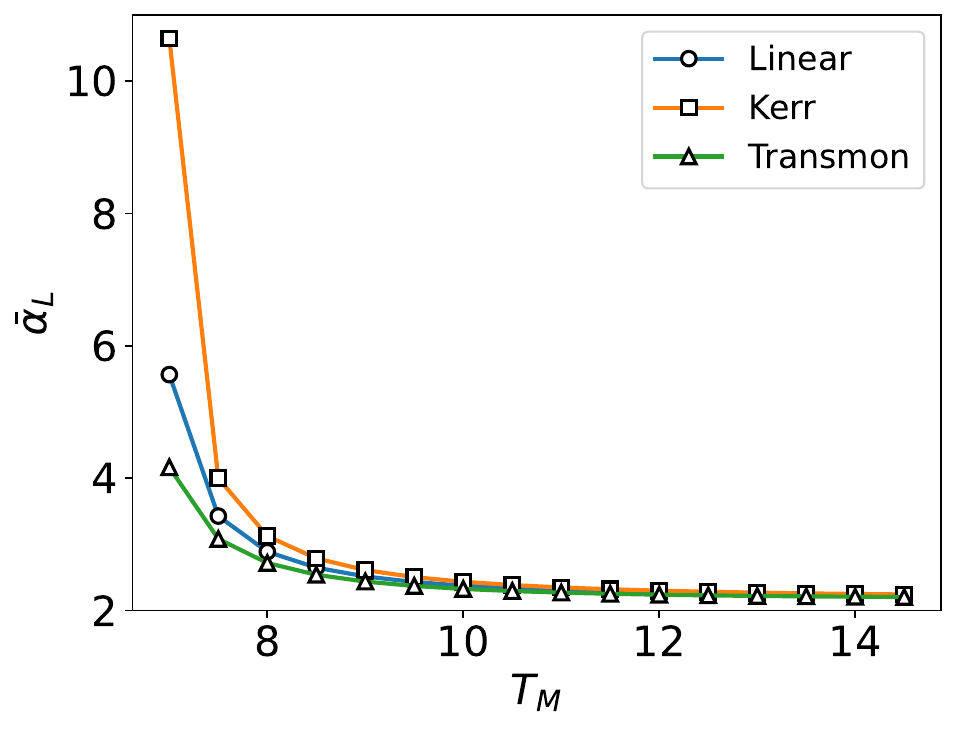}
    \caption{{ The figure depicts the dependence of $\bar{\alpha}_L$ on $T_M$ for linear, Kerr, and transmon environments, shown by the blue, orange, and green curves, respectively.}}
    \label{leb_12}
\end{figure}

\begin{figure*}[htbp]
    \subfigure[]{\includegraphics[width=0.4\textwidth]{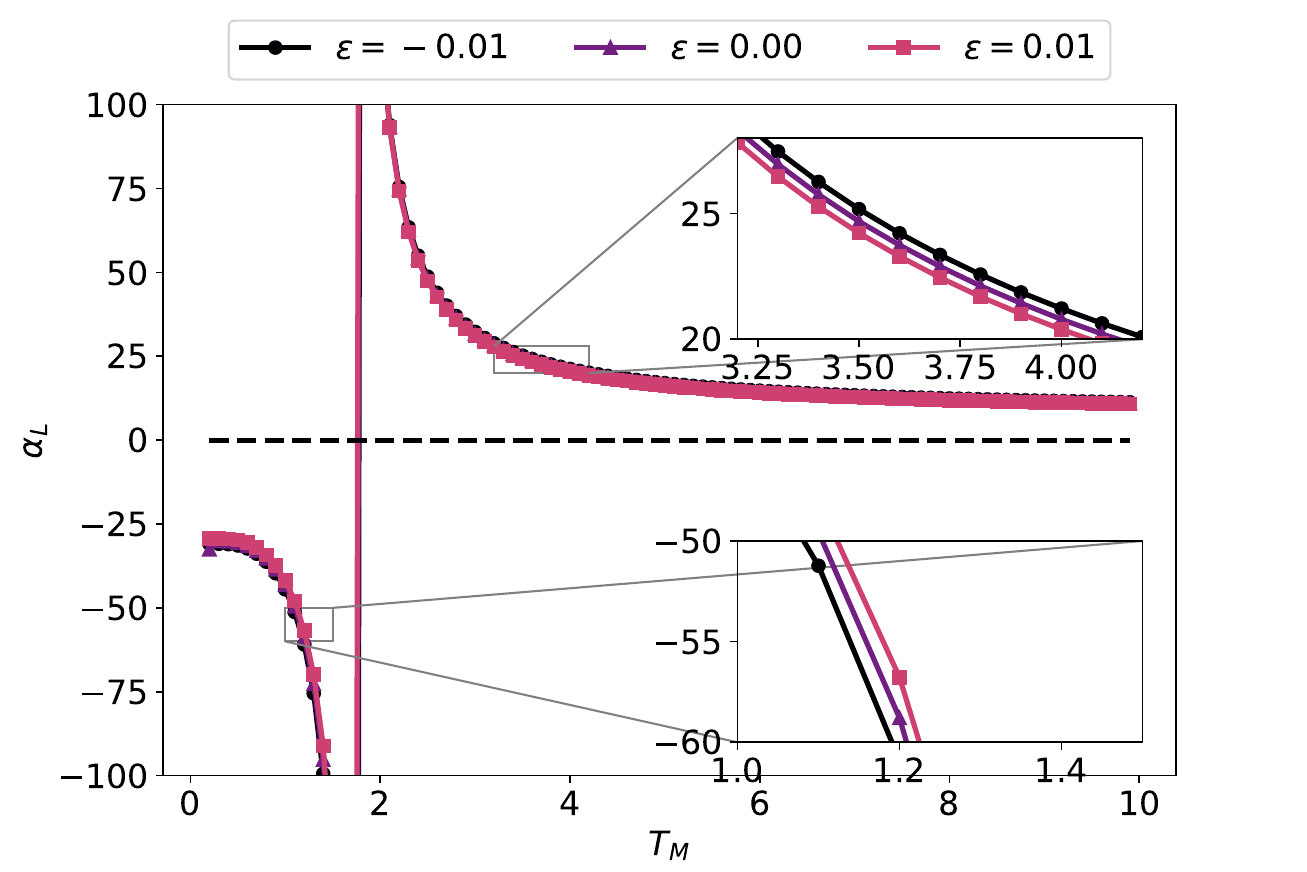}} 
    \label{a}
    \subfigure[]{\includegraphics[width=0.4\textwidth]{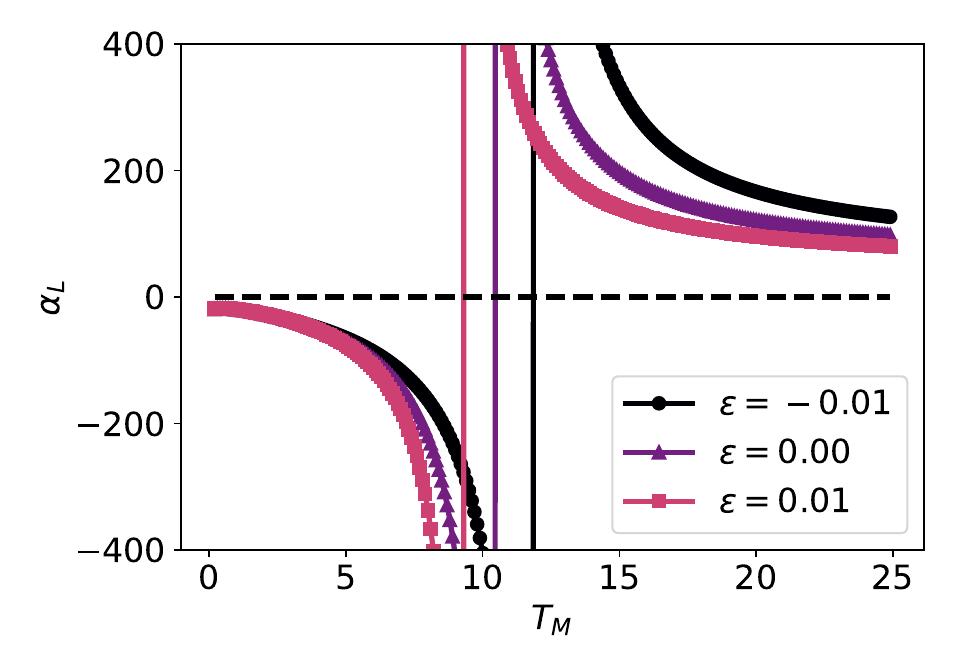}} 
    \label{b}
    \caption{Variation of amplification ($\alpha_L$) versus the temperature of the middle bath ($T_M$). The numerical analysis is done at time $t=0.4\tilde{t}$. Panel (a) corresponds to symmetric coupling between the qubits of the working substance i.e. $\omega_{ML} = \omega_{MR} = \omega_{LR} = \Delta$. The inset in the figure shows zoomed in area with $3.2<T_M<4.2$. Panel (b) corresponds to asymmetric coupling between the working substance qubits i.e. $\omega_{ML}= \Delta,\omega_{MR}= \Delta + \delta, \omega_{LR} = \Delta - \delta$, where $\delta=0.1$.}
    \label{symm-vs-asymm-nl}
\end{figure*}

\section{Response to non-linear environments in amplification of WTT}
\label{non-linear-sec}

Equispaced energy levels in a three-level system is a highly specific scenario and imposes strict constraints on the system, while the presence of non-linearity in systems is more probable in nature. Therefore, it is necessary to investigate the impact of non-linear environments on transistor action. Furthermore, there may arise frail perturbations in the energy levels of the environment qutrit, which generate non-linearity in the environment.
In this section we analyze the dynamical amplification factor of a WTT in presence of non-linear
environment. This means that the spacings between the energy levels are no longer equal as mentioned in Sec.~\ref{prelims}.

The local Hamiltonian of the environment connected to each qubit is given by 
\begin{equation}\label{8}
    H_{env}
=-\hbar
\begin{bmatrix}
\Delta & 0 & 0\\
0 & \epsilon & 0\\
0 & 0 & -\Delta\\
\end{bmatrix},
\end{equation}
where $\epsilon$ is indicative of the amount of non-linearity in the environment.
The condition, $\epsilon < 0$, corresponds to the three-level system being a  transmon qutrit, 
whose energy gaps gradually decrease as we go to higher energy levels.
The condition, $\epsilon > 0$, on the other hand, corresponds to Kerr type non-linearity of the environment-qutrit, where the energy gap increases as we consider higher energy levels. The quantity, $|\epsilon|$, can be considered to the strength of non-linearity (for either transmon or Kerr) present in the environment.

The three panels of Fig.~\ref{non-linearity-graph} depict the behavior of the dynamical amplification factor for the left qubit, $\alpha_L$, as a function of the temperature of the middle environment, $T_M$, for different values of the non-linearity strength, $|\epsilon|$.
In panel (a), the amplification for the linear case, transmon, and Kerr type environments are given by the purple, black and orange curves respectively. There is a discontinuity for all these three cases occurring at $T_M = T_M^{critical}$. The amplification, in each of the cases, increase as we increase $T_M$ when $T_M < T_M^{critical}$.
We further find that in the low temperature regime, i.e. to the left of the discontinuity, Kerr type non-linearity provides a better amplification than the linear case, while in high temperature regime, i.e. to the right of the discontinuity, transmon qutrits provide an advantage in terms of better amplification than the linear case. 
In Fig.~\ref{non-linearity-graph}-(b), the dynamical amplification factor is plotted for different values of strength of non-linearity, $|\epsilon|$, and for $\epsilon>0$, i.e. in the Kerr-type regime. Similar to panel (a), a discontinuity in the behavior of amplification is observed. 
At low $T_M$ limit, to the left of the discontinuity of the orange curve, corresponding to $\epsilon=0.05$, an increase in the value of non-linearity strength increases the amplification for a fixed $T_M$. However, to the right of the discontinuity corresponding to the value curve corresponding to $\epsilon=0.01$, the feature is just the opposite, i.e. as we increase the strength of non-linearity, the amplification also increases.

In Fig.~\ref{non-linearity-graph}-(c), the dynamical amplification factor is plotted for different values of $|\epsilon|$, corresponding to $\epsilon<0$, i.e. in the transmon regime. Similar to the previous scenarios, here also a discontinuity in the behavior of amplification is noted. To the left of the discontinuity of black curve corresponding to $\epsilon=-0.01$, an increase in the value of the non-linearity strength decreases the amplification with $T_M$. However, to the right of the discontinuity of the orange curve corresponding to $\epsilon=-0.05$, the behavior is exactly the opposite and increasing the strength of non-linearity causes an increase in amplification.

We further consider how the variation of amplification with time gets affected when non-linearity is introduced in the system, and demonstrate the results in   Fig.~\ref{amp-time-non-linear}.
For the transmon and Kerr type environments,
we observe that the amplification as a function of time, follows a trend which is similar 
to that of the linear environment case. 
However, the change in amplification when we consider non-linear environments is very small ($\sim 10^{-2}$) at all points of time except at time $t=\Tilde{t}$, where the amplification has significant differences, i.e. $\epsilon = -0.01$ gives an amplification of $38.98$, $\epsilon = 0.01$ gives an amplification of $34.00$, meanwhile linear type of environment gives an amplification of $36.27$. 
The maximum amplification that one can attain using this setup in presence of Kerr type environments is slightly higher $(38.98)$
than that using transmon qutrits $(34.00)$,
albeit an amplification close to the linear case is obtained in both the cases. 


{These repeated high-amplification peaks provide clear evidence of thermal transistor behavior in the device. But a characterization based solely on instantaneous amplification may not fully capture the overall performance of the transistor in the presence of periodic dynamics.}

{Here, we have introduced a cycle-averaged amplification factor, denoted by $\bar{\alpha}_L$, as an additional figure of merit. By averaging over a number of periodic cycles, $\bar{\alpha}_L$ reduces the influence of temporal oscillations and provides a more representative measure of the amplification achieved during the periodic steady state. While the instantaneous amplification factor highlights the occurrence of transistor action at specific operating points, the cycle-averaged quantity characterizes the overall performance of the device over an entire cycle. { The Hamiltonian parameters are identical to those used in the study of {Fig.~\ref{alpha-L-var-gk}}}. The average is evaluated over the time interval $t \in [0,5]$, which contains ten amplification peaks. In Fig.~\ref{leb_12}, we plot the time averaged value of $\alpha_L$ denoted as $\bar{\alpha}_L$ as a function of $T_M$ for three different environments: linear, Kerr, and transmon. The blue, orange, and green curves correspond to the linear, Kerr, and transmon environments, respectively. We find that the cycle-averaged amplification factor remains larger than 2 for all three cases, demonstrating that the transistor effect persists irrespective of the specific environmental nonlinearity. Moreover, in the temperature range $T_M \in [7,9]$, the Kerr environment yields the largest cycle-averaged amplification, outperforming both the linear and transmon environments. These results further confirm the robustness of the transistor effect when assessed through a cycle-averaged figure of merit.}



Next we introduce symmetric and asymmetric coupling between the qubits of the working substance along with non-linearities in the individual bath qutrits.  
The local Hamiltonian of the environment connected to each qubit is given by Eqn.~\ref{8} as discussed in the previous section. 
Meanwhile, the coupling between working-substance qubits are given by Eqns.~\ref{symm-couple} and~\ref{unsymm-couple} for symmetric and asymmetric coupling respectively. In particular, we have taken $\delta = 0.1$. The numerical values of amplification $\alpha_L$ are plotted versus the temperature of the middle bath $T_M$ for the symmetric and asymmetric cases in Fig.~\ref{symm-vs-asymm-nl} (a) and (b) respectively. 
We show that a critical temperature exists for both symmetric and asymmetric coupling, at which a discontinuity in amplification occurs. These critical points are denoted by $T^{\text{critical}}_{M,S}$ and $T^{\text{critical}}_{M,A}$ for symmetric and asymmetric coupling, respectively.
The insets in Fig.~\ref{symm-vs-asymm-nl} provides a magnified view of regions below and above the critical temperature. We find that in both such regions, transmon-type environment offers an enhanced amplification. This is particularly different in the previous set of coupling defined by Eq.~\ref{og-couple} where in transmon and Kerr both showed an advantage in different regions. However, in the asymmetric coupling case, the relative advantage of transmon- and Kerr-type environments follows the same qualitative behavior as observed earlier. Specifically, in the low-temperature regime (to the left of the discontinuity, i.e. $T_M<T^{critical}_{M,A}$), Kerr-type non-linearity yields stronger amplification compared to the linear case, while in the high-temperature regime (to the right of the discontinuity, i.e. $T_M<T^{critical}_{M,A}$), transmon qutrits provide superior amplification relative to the linear case.

\section{Conclusion}
\label{conclusions-sec}

Quantum thermal transistors with Markovian baths have been rigorously studied and worked upon. Non-Markovian systems, however, are 
often
more plausible in realistic situations. 
In this work, we considered a type of transistor that interacts with the environment via a collisional model consisting of a rail of qutrits, thereby deviating from Markovianity. The working substance consists of three qubits, i.e. left, right, and middle qubits. We found that such an apparatus, which we referred to as a working-substance thermal transistor, does manifest the transistor effect.  The interactions between the right-middle and the left-middle qubits are considered to be equal, but there is no interaction between the left and right qubits.
Such a nomenclature owes its origin to the fact that we focused on the heat currents which flow in and out of each qubit of the working substance 
due to the heat flow 
from and to the different parts of the system and environment, instead of on the currents which flow in and out of a single bath.
We depicted how the figure of merit of the transistor, i.e. the amplification of the heat currents, vary with the temperature of the modulating environment, system-environment coupling, and the interaction time. 
{We further demonstrate that the amplification remains robust under variations of the system-bath interaction strength, collision time, number of interactions, and initial-state preparations.}
We have also analyzed the variation of amplification when one of the baths, that is not  the modulating one, is removed. 

Furthermore, we studied the effects of frail non-linearities in the environment, which are either of transmon- or Kerr-type. We found a significant amplification in each of these cases, and showed that both the type of non-linearities prove to be beneficial in different temperature regimes.
In all of these cases, we also analyzed the variation of  amplification with time. 
We found a non-monotonicity in the behavior of amplification versus time. 
{We also examine the cycle-averaged amplification and find that it exhibits a clear transistor-like response for all three types of environments considered in this work.}

%
We further analyze two scenarios. In the first, the couplings among the left-middle, right-middle, and left-right qubit pairs are taken to be unequal, corresponding to the asymmetric case. In the second, all three couplings are chosen to be equal, representing the symmetric case. In both settings, we observe sufficient amplification for both the left and right qubits of the working substance. We also demonstrate a pronounced transistor effect for the left and right qubits of the working substance when the environment consists of qubits, and separately when the environment  pertains non-linearities, following the same collisional model. 
\\
\vspace{-25pt}
\acknowledgements 
We acknowledge computations performed using Armadillo \cite{arma1,arma2} on the cluster computing facility of Harish-Chandra Research Institute (HRI), India.
  We acknowledge the use of \href{https://github.com/titaschanda/QIClib}{QIClib} - a modern C++ library for general purpose quantum information processing and quantum computing.
 %
 DB thanks HRI for support during visits.  AB acknowledges support from ‘INFOSYS scholarship for senior students’ at Harish Chandra Research Institute, India. US acknowledges financial support from the Anusandhan National Research Foundation (ANRF), Government of India, under the Grant No. ANRF/ARG/2025/004617/PS.
\onecolumngrid
\section*{Appendix}
\appendix
\setcounter{figure}{0}
\section{Amplification with two qubits and two environments} \label{appA}
In this subsection, we consider the case where the system comprises two qubits, left ($L$) and right ($R$), which are connected to respective environments having temperatures $T_L$ and $T_R$ (in dimensions of $\Tilde{T}$) respectively. In this situation, we consider the left bath to be the modulating bath and see how a change in temperature of 
the left bath affects the amplification of the current in the right qubit. The dynamical amplification factor, in this case, is defined as
\begin{equation}\nonumber
       \alpha = \frac{\partial J_R}{\partial J_L} = \frac{\frac{\partial J_R}{\partial T_M}}{\frac{\partial J_L}{\partial T_M}},
\end{equation}
where $J_L$ and $J_R$ are the heat currents flowing into the left and right qubits respectively.
In this case, we observe that when the energy difference between the ground state and first excited state is the same for left and right qubit, then there is no amplification.
However, when we consider unequal energy differences, there is a finite amplification in a small range of $T_L$. The range of $T_L$ is from 0 to 2.488 when $\omega_L = 1$ and $\omega_R = 2$, $T_R = 4, g=4, \Delta=5$ and $t=1$. 
In conventional classical transistors, the three components, i.e. emitter, base and collector, are all essential to observe the transistor effect, and removing any one of them suppresses this behavior. Prototypically, a two-terminal electronic device that demonstrates transistor-like effects cannot be regarded as a transistor in the classical sense. However, in the quantum mechanical domain, a two-terminal device that exhibits properties similar to a transistor can be referred to as a quantum transistor. In our work, we observe such a transistor effect in a specific regime of the left-bath temperature, $T_L$, when the energy gaps of the first and second excited states differ between the left and right qubits. This effect is purely quantum in nature, being absent in the corresponding classical counterparts.

\setcounter{figure}{0}
\section{Environment with Qubits}
\label{qubit_env}
In this section, we aim to explore what happens when the environment consists of qubits instead of qutrits. Our primary objective is to determine whether a transistor effect persists when the working substance interacts with this type of environment. 

Here we provide an analysis of the transistor action in a similar WTT model but with the environmental qutrits being replaced by qubits. In this regard, the local Hamiltonian of a single qubit of the environment gets modified to
\begin{equation}
    H_{env} = -\hbar~ \Delta \sum_{i\in\{L,M,R\}} \sigma_z^{(i)},
\label{bath-eqn-qubit}
\end{equation}
where $\sigma_z^{(i)}$ is the Pauli-$z$ matrix for spin-1/2 particles. Similarly, the interaction Hamiltonian also gets modified to
\begin{equation}
    H_{qubit-env} = - \hbar~g\sum_{i\in\{L,M,R\}}\sigma_x^{(i),qubit}~\sigma_x^{(i),qubit},
\end{equation}
%
where $\sigma_x^{(i),qubit}$ is the Pauli-$x$ matrix for spin-$1/2$ particle, and $g$ is the coupling constant between the system-qubit and the environment-qubit. The parameters considered are: $T_L = 4,~ T_R = 10, ~g=4, ~\Delta=3, ~T=5$. 

\begin{figure}[htbp]
    \centering
    \subfigure[]{\includegraphics[width=0.4\textwidth]{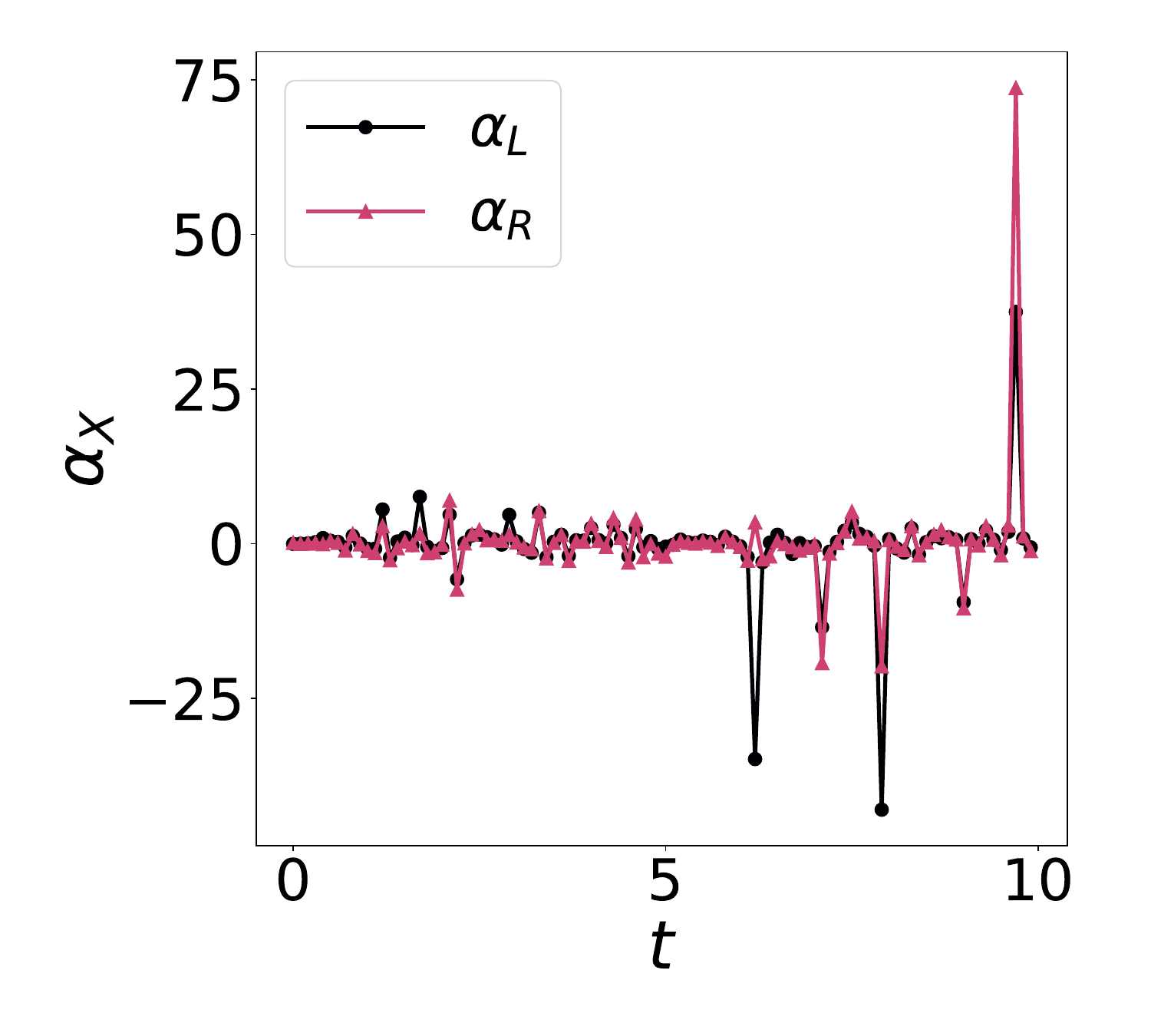}}
    \label{popx}
    \subfigure[]{\includegraphics[width=0.4\textwidth]{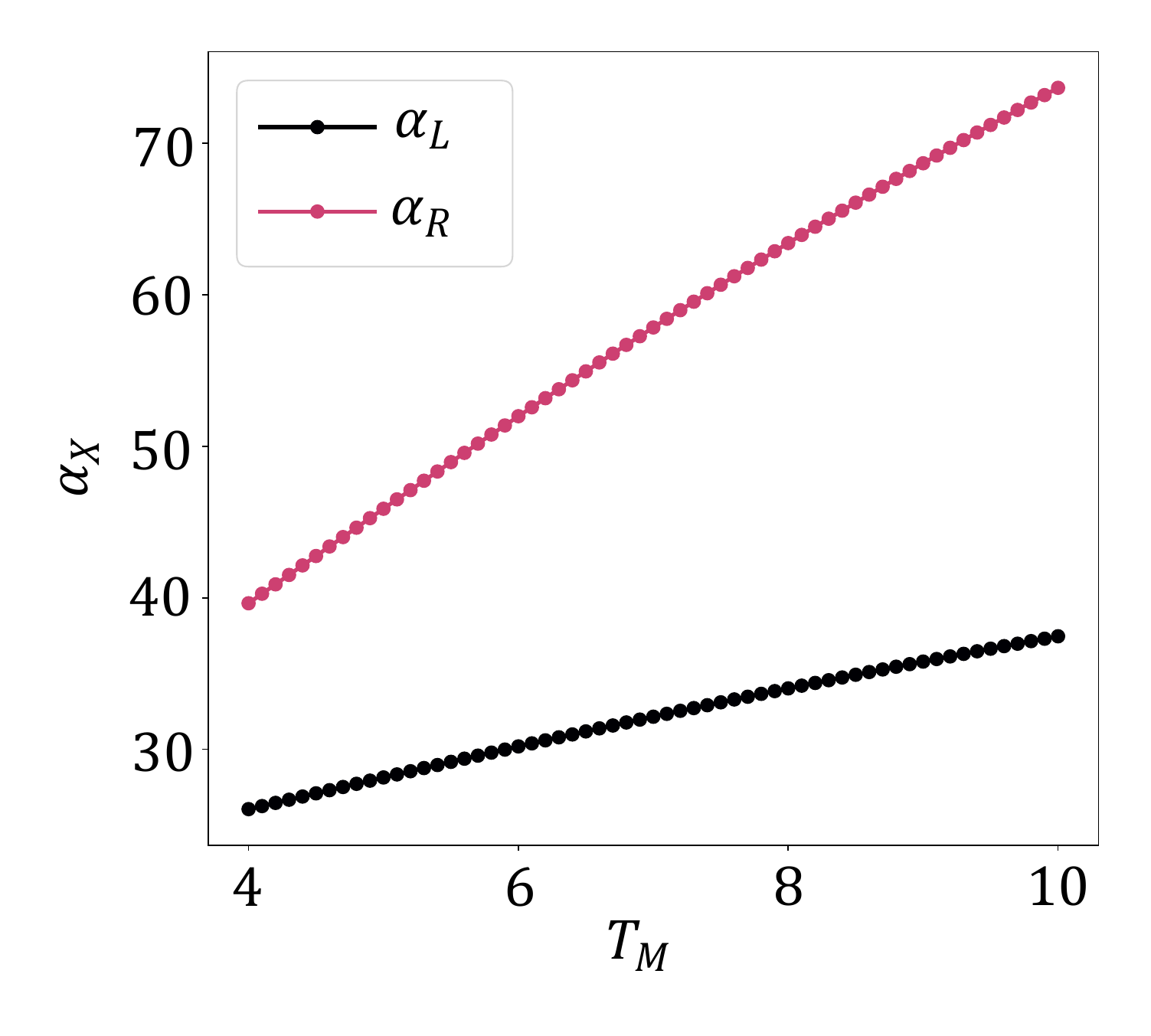}}
    \label{1x1b}
    \caption{In panel (a) Variation of the dynamical amplification factor, $\alpha_L$ and $\alpha_R$, with time, $t$. The vertical axis is dimensionless and the horizontal axis is in units of $\tilde{t}$. The given graph is at $T_M = 10$ in units of $\Tilde{T}$. At the same time in panel (b) Variation of the dynamical amplification factor, $\alpha_L$, along the vertical axis, versus the temperature of the middle bath, $T_M$, along the horizontal axis. The quantity, $\alpha_L$, is dimensionless, while $T_M$ is in units of $\tilde{T}$. The graph is plotted at time $t= 9.7$.}
    \label{non-linearity-graphx}
\end{figure}

We have plotted the dynamic curves for the dynamical amplification factors, $\alpha_L$ and $\alpha_R$, as functions of time $t$ in Fig.~\ref{non-linearity-graphx} (a). The maximum values of amplification for the left and right qubits, $\alpha_L$ and $\alpha_R$, are observed at time $9.7 \tilde{t}$, with the corresponding maximum values being 37.46 and 73.67.
At the same time, we have examined how the amplification associated with the left and right qubits changes as the temperature of the middle bath is varied. This behavior is illustrated in Fig.~\ref{non-linearity-graphx} (b), where it is shown that the amplifications for both qubits increase monotonically with an increase in the  temperature of the middle bath.
Hence, a two-level system will also be sufficient constituent of a collisional model as environment to observe transistor action. \\




\section{{Energy conservation}}\label{app_cc}

{The condition
$\mathcal{J}_L+\mathcal{J}_M+\mathcal{J}_R=0$
usually arises in studies of thermal transistors operating in a Markovian steady-state regime. In such treatments, the quantities $\mathcal{J}_X$ ($X\in{L,M,R}$) represent the \emph{global heat currents} flowing between the system and the corresponding thermal reservoirs. 
For clarity, we first briefly review the standard derivation of the global heat currents in a Markovian master-equation framework. This derivation is included for comparison with the local-heat current approach  discussed below Eq.~\eqref{c2}.

Starting from a master equation of the form}
{\begin{equation}\label{C1}
\frac{d\rho}{dt}
=
-i[H_s,\rho]
+\sum_{P=L,M,R}\mathcal{L} P[\rho],
\end{equation}}
{the system energy is given by
$E=\mathrm{Tr}(H_s\rho)$.
Taking the time derivative yields}
{\begin{equation}\label{C2}
\frac{dE}{dt}
=
\mathrm{Tr}\left(H_s\frac{d\rho}{dt}\right)
\nonumber\
=
-i\mathrm{Tr}\left(H_s[H_s\rho]\right)
+
\sum{P=L,M,R}
\mathrm{Tr}\left(H_s\mathcal{L}P[\rho]\right)
\nonumber\
=
\sum{P=L,M,R}
\mathrm{Tr}\left(H_s\mathcal{L}_P[\rho]\right),
\end{equation}}
{where we have used
$\mathrm{Tr}\left(H_s[H_s,\rho]\right)=0$.
Identifying the bath heat currents as
$\mathcal{J}_P=
\mathrm{Tr}\left(H_s\mathcal{L}_P[\rho]\right)$,
one obtains}
{\begin{equation}
\frac{dE}{dt}
=
\mathcal{J}_L+\mathcal{J}_M+\mathcal{J}_R.
\label{c2}
\end{equation}}
{In the steady state, $dE/dt=0$, leading to the familiar condition
$\mathcal{J}_L+\mathcal{J}_M+\mathcal{J}_R=0$.}

The above   equations \eqref{C1} and \eqref{C2} describe the \emph{global heat currents} within a Markovian master-equation framework. In contrast, the currents considered in our work are \emph{local heat currents}, whose derivation follows from the reduced dynamics generated by the collision model. In this model, the system interacts sequentially with environmental qutrits, and the resulting reduced dynamics is described by a  of completely positive and trace-preserving (CPTP) maps rather than by a Markovian master equation. 
If the initial state of the system and environment are $\rho_0$ and $\rho_E$, then the time derivative of the reduced density matrix of $X$'th working substance is given by $\dot{\rho}_X(t)=\mathrm{Tr}_{\bar{X}}\!\left[\dot{\rho}_S(t)\right]$, where $\rho_S(t)=\mathrm{Tr}_E\!\left[U(t)\left(\rho_S(0)\otimes\rho_E\right)U^\dagger(t)\right]$ for ($t<\delta t$) and $U(t)=e^{-iHt}$. For $t>\delta t$, say, the $n$'th collision $\dot{\rho}_X$ can be evaluated the same way with the global evolution of the $U(t)\rho_{n-1} \otimes \rho_{env}U^{\dag}(t)$ followed by tracing out the environment and $\bar{X}$.
The quantities appearing in Eq.~\eqref{heat-current-def} are local energy currents defined by
{\begin{equation}
J_X=\mathrm{Tr}\left(\dot{\rho}_X H_X\right),
\qquad
X\in{L,M,R},
\end{equation}}
{where $\rho_X$ and $H_X$ denote the reduced state and local Hamiltonian of subsystem $X$, respectively.  
For the full three-qubit system, the total Hamiltonian is}
{\begin{equation}
H_{\mathrm{sys}}
=
H_L+H_M+H_R+H_I,
\end{equation}}
{where $H_I$ denotes the interaction Hamiltonian between the qubits. The rate of change of the total system energy is therefore}
{\begin{equation}
\frac{dE}{dt}
=
\mathrm{Tr}\left(\dot{\rho}H_{\mathrm{sys}}\right)
=
\mathrm{Tr}(\dot{\rho}_L H_L)
+
\mathrm{Tr}(\dot{\rho}_M H_M)
+
\mathrm{Tr}(\dot{\rho}_R H_R)
+
\mathrm{Tr}(\dot{\rho}H_I)
=
J_L+J_M+J_R+J_I,
\end{equation}}
{where
$J_I=\mathrm{Tr}(\dot{\rho}H_I)$
accounts for the rate of change of the interaction energy. Therefore, the local currents do not, in general, satisfy the relation
$J_L+J_M+J_R=0$.
Even if the dynamics were to reach a stationary regime, the interaction-energy contribution must be taken into account when discussing energy conservation. Consequently, the condition satisfied by the global heat currents in conventional steady-state transistor models is not expected to hold for the local currents considered in our work.}

{In collision models, the repeated coupling and decoupling of the environment may be associated with switching work. Such contributions are absent in the idealized steady-state Markovian description but can arise naturally in collision-based dynamics. In our case, the dynamics is described by a general CPTP map, and the energy-balance relation is therefore modified by an additional contribution arising from the interaction energy between the qubits. Consequently, the rate of change of the total energy is given by}
{\begin{equation}
\frac{dE}{dt}=J_L+J_M+J_R+J_I,
\end{equation}}

{where
$J_I=\mathrm{Tr}(\dot{\rho}H_I)$
accounts for the interaction-energy contribution.  If the dynamics were to reach a stationary regime and no additional work contributions were present, one would expect
$J_L+J_M+J_R+J_I=0$.  Therefore, the local currents do not, in general, satisfy the relation
$J_L+J_M+J_R=0$.
Even if the dynamics were to reach a stationary regime, the interaction-energy contribution must be taken into account when discussing energy conservation. }

{This modification of the energy conservation equation arises naturally in general CPTP dynamics beyond the conventional Markovian framework.
In realistic thermal devices operating under general open-system interactions,  such inherent energy flows are  unavoidable. Moreover, it is to note that the amplification factor is defined entirely in terms of the local currents and therefore remains well defined irrespective of whether the system reaches a steady state. Indeed, in many quantum thermal devices, such as thermal transistors and quantum refrigerators, physically relevant phenomena including amplification and cooling are often studied in the transient regime, where the system has not yet attained its steady state~\cite{Transiant_transistor}. Even for Markovian systems, the condition
$J_L+J_M+J_R=0$
is generally satisfied only in the steady state and need not hold during transient evolution.}

\section{{Transistor effect with varied initial states}}\label{app_init_states}
\begin{figure}
    \centering
    \includegraphics[width=0.49\linewidth]{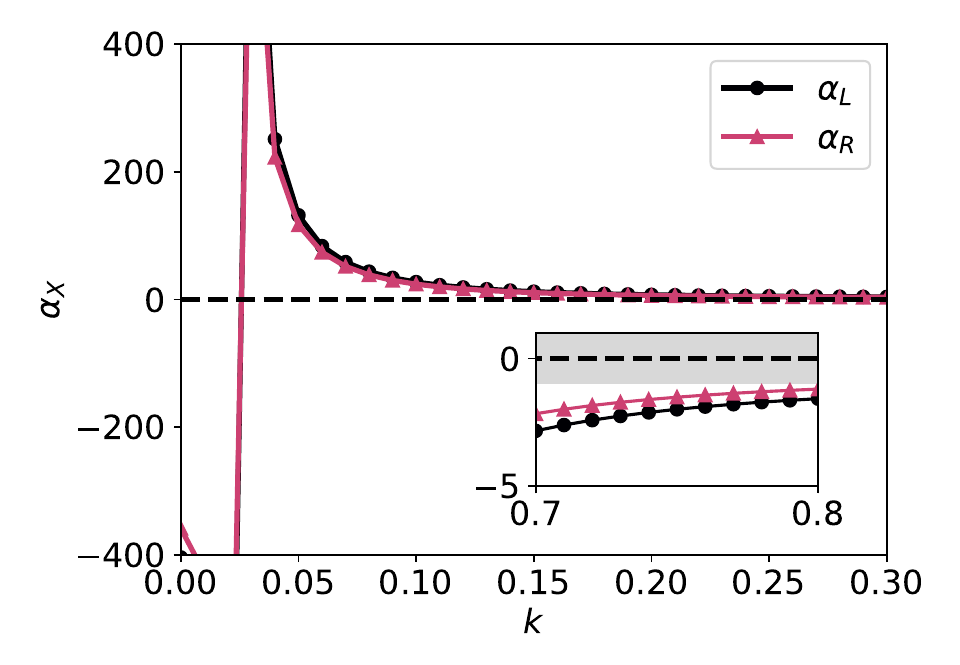}
    \includegraphics[width=0.49\linewidth]{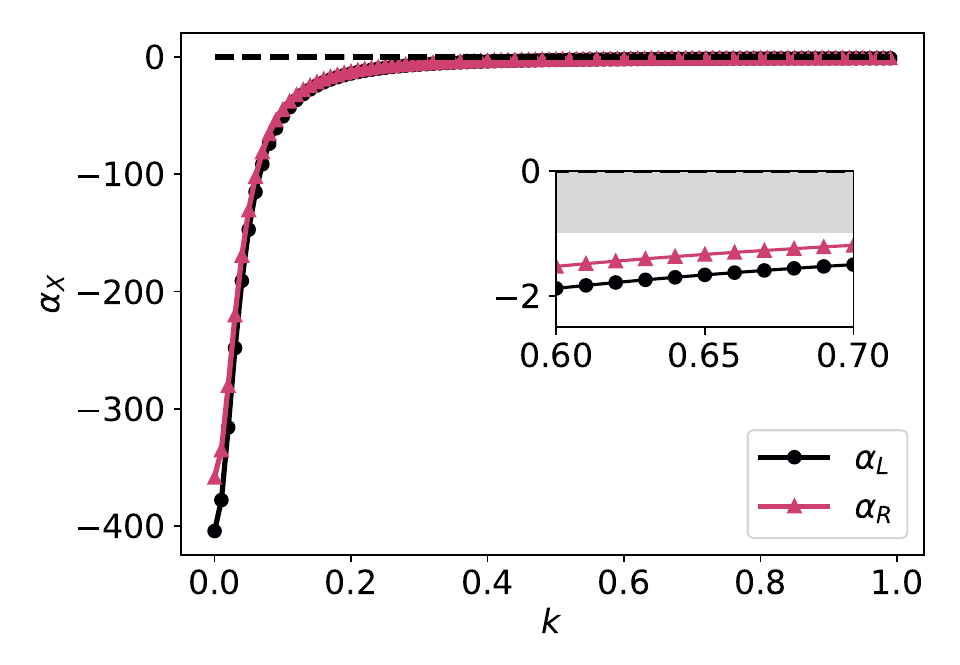}
    \caption{{Amplification for different initial system states. (Left) shows the amplification versus $k$ when the initial state of the system is taken to be $(\sin(k\pi/2)\ket{0} +\cos(k\pi/2)\ket{1})^{\otimes 3} $. As we vary the parameter $k$, the system state goes from $\ket{000}$ to $\ket{+++}$ and then finally to $\ket{111}$ (Right) shows the amplification versus $k$ when the initial state of the system is taken to be $\sin(k\pi/2)\ket{000} + \cos(k\pi/2)\ket{111}$. As we vary the parameter $k$, the state goes from $\ket{000}$ to $\ket{111}$.}}
    \label{fig:diff-init-states}
\end{figure}

{In addition to the initial system state $\ket{0}^{\otimes 3}$ considered in the main text, we here investigate two additional classes of initial states. The first class is $\ket{\psi}^{coh}_{in}=(\sin(k\pi/2)\ket{0} +\cos(k\pi/2)\ket{1})^{\otimes 3}$, for $k \in [0,1]$. Unlike $\ket{0}^{\otimes3}$, this class of states generally possesses a non-zero amount of coherence. In the left panel of Fig.~\ref{fig:diff-init-states}, we show that prominent transistor effects are observed over a broad range of $k \in [0,0.8]$.
The second class of states is $\ket{\psi}^{GHZ}_{in}=\sin(k\pi/2)\ket{000} + \cos(k\pi/2)\ket{111}$, for $k \in [0,1]$, which corresponds to the well-known family of generalized GHZ states. In the right panel of Fig.~\ref{fig:diff-init-states}, we demonstrate that prominent transistor action persists over a broad range of $k \in [0,0.7]$.}


\bibliography{transistor.bib}

@article{qtt-first-paper,
  title = {Quantum Thermal Transistor},
  author = {Joulain, K. and Drevillon, J. and Ezzahri, Y. and Ordonez-Miranda, J.},
  journal = {Phys. Rev. Lett.},
  volume = {116},
  issue = {20},
  pages = {200601},
  numpages = {5},
  year = {2016},
  month = {May},
  publisher = {American Physical Society},
  doi = {10.1103/PhysRevLett.116.200601}
}

@article{floquet_tt,
  title = {Floquet quantum thermal transistor},
  author = {Gupt, N. and Bhattacharyya, S. and Das, B. and Datta, S. and Mukherjee, V. and Ghosh, A.},
  journal = {Phys. Rev. E},
  volume = {106},
  issue = {2},
  pages = {024110},
  numpages = {14},
  year = {2022},
  month = {Aug},
  publisher = {American Physical Society},
  doi = {10.1103/PhysRevE.106.024110},
  url = {https://link.aps.org/doi/10.1103/PhysRevE.106.024110}
}

@article{qubit-qutrit,
  title = {Quantum thermal transistor based on qubit-qutrit coupling},
  author = {Guo, B. and Liu, T. and Yu, C.},
  journal = {Phys. Rev. E},
  volume = {98},
  issue = {2},
  pages = {022118},
  numpages = {8},
  year = {2018},
  month = {Aug},
  publisher = {American Physical Society},
  doi = {10.1103/PhysRevE.98.022118},
  url = {https://link.aps.org/doi/10.1103/PhysRevE.98.022118}
}

@article{exp_magnetic,
  title = {A three-terminal magnetic thermal transistor},
  volume = {14},
  ISSN = {2041-1723},
  url = {http://dx.doi.org/10.1038/s41467-023-36056-4},
  number = {1},
  journal = {Nature Communications},
  publisher = {Springer Science and Business Media LLC},
  author = {Castelli,  L. and Zhu,  Q. and Shimokusu,  T. J. and Wehmeyer,  G.},
  year = {2023},
  month = jan 
}

@article{exp-super-1,
  title = {Photonic heat transport in three terminal superconducting circuit},
  volume = {13},
  ISSN = {2041-1723},
  url = {http://dx.doi.org/10.1038/s41467-022-29078-x},
  number = {1},
  journal = {Nature Communications},
  publisher = {Springer Science and Business Media LLC},
  author = {Gubaydullin,  A. and Thomas,  G. and Golubev,  D. S. and Lvov,  D. and Peltonen,  J. T. and Pekola,  J. P.},
  year = {2022},
  month = mar 
}

@article{exp-super-2,
  title = {Quantum thermal transistor in superconducting circuits},
  author = {Majland, M. and Christensen, K. Sangild and Z., Nikolaj T.},
  journal = {Phys. Rev. B},
  volume = {101},
  issue = {18},
  pages = {184510},
  numpages = {10},
  year = {2020},
  month = {May},
  publisher = {American Physical Society},
  doi = {10.1103/PhysRevB.101.184510},
  url = {https://link.aps.org/doi/10.1103/PhysRevB.101.184510}
}

@article{exp-vo2,
  title = {Transistorlike Device for Heating and Cooling Based on the Thermal Hysteresis of ${\mathrm{VO}}_{2}$},
  author = {Ordonez-Miranda, J. and Ezzahri, Y. and Drevillon, J. and Joulain, K.},
  journal = {Phys. Rev. Appl.},
  volume = {6},
  issue = {5},
  pages = {054003},
  numpages = {7},
  year = {2016},
  month = {Nov},
  publisher = {American Physical Society},
  doi = {10.1103/PhysRevApplied.6.054003},
  url = {https://link.aps.org/doi/10.1103/PhysRevApplied.6.054003}
}

@article{exp-vo2-2,
  title = {VO 2 -based radiative thermal transistor in the static regime},
  author = {Prod'Homme,  H. and Ordonez-Miranda,  J. and Ezzahri,  Y. and Drevillon,  J. and Joulain,  K.},
   journal = {arXiv:1710.10332},
year = {2017},
  url = {https://arxiv.org/abs/1710.10332},
}

@article{heat-current-1,
  title = {Markovian master equations for quantum thermal machines: local versus global approach},
  volume = {19},
  ISSN = {1367-2630},
  url = {http://dx.doi.org/10.1088/1367-2630/aa964f},
  DOI = {10.1088/1367-2630/aa964f},
  number = {12},
  journal = {New Journal of Physics},
  publisher = {IOP Publishing},
  author = {Hofer,  P. P and Perarnau-Llobet,  M. and Miranda,  L. D. M. and Haack,  G. and Silva,  R. and Brask,  J. B. and Brunner,  N.},
  year = {2017},
  month = dec,
  pages = {123037}
}

@article{heat-current-2,
  title = {Quantum thermodynamically consistent local master equations},
  volume = {3},
  ISSN = {2643-1564},
  url = {http://dx.doi.org/10.1103/PhysRevResearch.3.013165},
  number = {1},
  journal = {Physical Review Research},
  publisher = {American Physical Society (APS)},
  author = {Hewgill,  A. and De Chiara,  G. and Imparato,  A.},
  year = {2021},
  month = feb 
}

@article{darlington-1,
  title = {Darlington pair of quantum thermal transistors},
  volume = {104},
  ISSN = {2469-9969},
  url = {http://dx.doi.org/10.1103/PhysRevB.104.045405},
  number = {4},
  journal = {Physical Review B},
  publisher = {American Physical Society (APS)},
  author = {Wijesekara,  R. T. and Gunapala,  S. D. and Premaratne,  M.},
  year = {2021},
  month = jul 
}

@article{darlington-energy-divider,
  title = {Towards quantum thermal multi-transistor systems: Energy divider formalism},
  volume = {105},
  ISSN = {2469-9969},
  url = {http://dx.doi.org/10.1103/PhysRevB.105.235412},
  number = {23},
  journal = {Physical Review B},
  publisher = {American Physical Society (APS)},
  author = {Wijesekara,  R. T. and Gunapala,  S. D. and Premaratne,  M.},
  year = {2022},
  month = jun 
}

@article{transistor-network,
  title = {Electrothermal Transistor Effect and Cyclic Electronic Currents in Multithermal Charge Transfer Networks},
  volume = {118},
  ISSN = {1079-7114},
  url = {http://dx.doi.org/10.1103/PhysRevLett.118.207201},
  number = {20},
  journal = {Physical Review Letters},
  publisher = {American Physical Society (APS)},
  author = {Craven,  G. T. and Nitzan,  A.},
  year = {2017},
  month = may 
}

@article{noise-1,
  title = {Stochastic model of noise for a quantum thermal transistor},
  volume = {108},
  ISSN = {2469-9969},
  url = {http://dx.doi.org/10.1103/PhysRevB.108.235421},
  number = {23},
  journal = {Physical Review B},
  publisher = {American Physical Society (APS)},
  author = {Ekanayake,  U. N. and Gunapala,  S. D. and Premaratne,  M.},
  year = {2023},
  month = dec 
}

@article{noise-2,
  title = {Common Environmental Effects on Quantum Thermal Transistor},
  volume = {24},
  ISSN = {1099-4300},
  url = {http://dx.doi.org/10.3390/e24010032},
  DOI = {10.3390/e24010032},
  number = {1},
  journal = {Entropy},
  publisher = {MDPI AG},
  author = {Liu,  Y. and Yu,  D. and Yu,  C.},
  year = {2021},
  month = dec,
  pages = {32}
}

@article{noise-3,
  title = {Engineered common environmental effects on multitransistor systems},
  author = {Ekanayake, U. N. and Gunapala, S. D. and Premaratne, M.},
  journal = {Phys. Rev. B},
  volume = {107},
  issue = {7},
  pages = {075440},
  numpages = {20},
  year = {2023},
  month = {Feb},
  publisher = {American Physical Society},
  doi = {10.1103/PhysRevB.107.075440},
  url = {https://link.aps.org/doi/10.1103/PhysRevB.107.075440}
}

@Article{noise-4,
AUTHOR = {Mandarino, A.},
TITLE = {Quantum Thermal Amplifiers with Engineered Dissipation},
JOURNAL = {Entropy},
VOLUME = {24},
YEAR = {2022},
NUMBER = {8},
ARTICLE-NUMBER = {1031},
URL = {https://www.mdpi.com/1099-4300/24/8/1031},
PubMedID = {35893011},
ISSN = {1099-4300}
}

@article{gen-1,
  title = {Multifunctional quantum thermal device utilizing three qubits},
  author = {Guo, B. and Liu, T. and Yu, C.},
  journal = {Phys. Rev. E},
  volume = {99},
  issue = {3},
  pages = {032112},
  numpages = {8},
  year = {2019},
  month = {Mar},
  publisher = {American Physical Society},
  doi = {10.1103/PhysRevE.99.032112},
  url = {https://link.aps.org/doi/10.1103/PhysRevE.99.032112}
}

@article{gen-2,
  title = {Minimal quantum heat manager boosted by bath spectral filtering},
  author = {Naseem, M. T. and Misra, A. and M\"ustecaplio\ifmmode \breve{g}\else \u{g}\fi{}lu, O. E. and Kurizki, G.},
  journal = {Phys. Rev. Res.},
  volume = {2},
  issue = {3},
  pages = {033285},
  numpages = {12},
  year = {2020},
  month = {Aug},
  publisher = {American Physical Society},
  doi = {10.1103/PhysRevResearch.2.033285},
  url = {https://link.aps.org/doi/10.1103/PhysRevResearch.2.033285}
}

@article{gen-3,
  title = {Quantum thermal transistors: Operation characteristics in steady state versus transient regimes},
  author = {Ghosh, R. and Ghoshal, A. and Sen, U.},
  journal = {Phys. Rev. A},
  volume = {103},
  issue = {5},
  pages = {052613},
  numpages = {14},
  year = {2021},
  month = {May},
  publisher = {American Physical Society},
  doi = {10.1103/PhysRevA.103.052613},
  url = {https://link.aps.org/doi/10.1103/PhysRevA.103.052613}
}

@article{gen-4,
  title = {Thermal Transistor Effect in Quantum Systems},
  author = {Mandarino, A. and Joulain, K. and G\'omez, M. Dom\'{\i}nguez and Bellomo, B.},
  journal = {Phys. Rev. Appl.},
  volume = {16},
  issue = {3},
  pages = {034026},
  numpages = {15},
  year = {2021},
  month = {Sep},
  publisher = {American Physical Society},
  doi = {10.1103/PhysRevApplied.16.034026},
  url = {https://link.aps.org/doi/10.1103/PhysRevApplied.16.034026}
}

@article{gen-5,
  title = {A multifunctional quantum thermal device: With and without inner coupling},
  volume = {393},
  ISSN = {0375-9601},
  DOI = {10.1016/j.physleta.2021.127172},
  journal = {Physics Letters A},
  publisher = {Elsevier BV},
  author = {Huangfu,  Y. and Qi,  S. and Jing,  J.},
  year = {2021},
  month = mar,
  pages = {127172}
}

@article{gen-6,
  title = {Dynamic response of a thermal transistor to time-varying signals},
  volume = {33},
  ISSN = {2058-3834},
  DOI = {10.1088/1674-1056/ad2dcc},
  number = {5},
  journal = {Chinese Physics B},
  publisher = {IOP Publishing},
  author = {Ruan ,  Q.  and Liu ,  W.  and Wang ,  L. },
  year = {2024},
  month = may,
  pages = {056301}
}

@article{gen-7,
  title = {Quantum thermal transistor: a unified method from weak to strong internal coupling},
  volume = {53},
  ISSN = {1361-6455},
  DOI = {10.1088/1361-6455/abade1},
  number = {20},
  journal = {Journal of Physics B: Atomic,  Molecular and Optical Physics},
  publisher = {IOP Publishing},
  author = {Yang,  H. and Tan,  Y.},
  year = {2020},
  month = sep,
  pages = {205504}
}

@article{columb-coupled-1,
  title = {Thermal transistor and thermometer based on Coulomb-coupled conductors},
  author = {Yang, J. and Elouard, C. and Splettstoesser, J. and Sothmann, B. and S\'anchez, R. and Jordan, A. N.},
  journal = {Phys. Rev. B},
  volume = {100},
  issue = {4},
  pages = {045418},
  numpages = {10},
  year = {2019},
  month = {Jul},
  publisher = {American Physical Society},
  doi = {10.1103/PhysRevB.100.045418},
  url = {https://link.aps.org/doi/10.1103/PhysRevB.100.045418}
}

@article{columb-coupled-2,
  title = {Coulomb-coupled quantum-dot thermal transistors},
  volume = {122},
  ISSN = {1286-4854},
  url = {http://dx.doi.org/10.1209/0295-5075/122/17002},
  DOI = {10.1209/0295-5075/122/17002},
  number = {1},
  journal = {EPL (Europhysics Letters)},
  publisher = {IOP Publishing},
  author = {Zhang,  Y. and Yang,  Z. and Zhang,  X. and Lin,  B. and Lin,  G. and Chen,  J.},
  year = {2018},
  month = apr,
  pages = {17002}
}

@article{polaron,
  title = {A polaron theory of quantum thermal transistor in nonequilibrium three-level systems*},
  volume = {29},
  ISSN = {1674-1056},
  url = {http://dx.doi.org/10.1088/1674-1056/ab973b},
  DOI = {10.1088/1674-1056/ab973b},
  number = {8},
  journal = {Chinese Physics B},
  publisher = {IOP Publishing},
  author = {Wang,  C. and Xu,  D.},
  year = {2020},
  month = jul,
  pages = {080504}
}

@article{cavity-optomechanical,
  title = {Quantum transistor with a double-cavity optomechanical system},
  volume = {122},
  ISSN = {1286-4854},
  url = {http://dx.doi.org/10.1209/0295-5075/122/64002},
  DOI = {10.1209/0295-5075/122/64002},
  number = {6},
  journal = {EPL (Europhysics Letters)},
  publisher = {IOP Publishing},
  author = {Xiong,  B. and Li,  X. and Chao,  S. and Zhou,  L.},
  year = {2018},
  month = jul,
  pages = {64002}
}

@article{normal-superconductor,
  title = {Three-terminal normal-superconductor junction as thermal transistor},
  volume = {92},
  ISSN = {1434-6036},
  url = {http://dx.doi.org/10.1140/epjb/e2019-90747-0},
  number = {2},
  journal = {The European Physical Journal B},
  publisher = {Springer Science and Business Media LLC},
  author = {Tang,  G. and Peng,  J. and Wang,  J.},
  year = {2019},
  month = feb 
}

@article{coherence-enhanced,
  title = {Coherence-enhanced thermal amplification for small systems},
  volume = {569},
  ISSN = {0378-4371},
  url = {http://dx.doi.org/10.1016/j.physa.2021.125753},
  DOI = {10.1016/j.physa.2021.125753},
  journal = {Physica A: Statistical Mechanics and its Applications},
  publisher = {Elsevier BV},
  author = {Su,  S. and Zhang,  Y. and Andresen,  B. and Chen,  J.},
  year = {2021},
  month = may,
  pages = {125753}
}

@article{ndtc-1,
  title = {Heat amplification and negative differential thermal conductance in a strongly coupled nonequilibrium spin-boson system},
  volume = {97},
  ISSN = {2469-9934},
  url = {http://dx.doi.org/10.1103/PhysRevA.97.052112},
  number = {5},
  journal = {Physical Review A},
  publisher = {American Physical Society (APS)},
  author = {Wang,  C. and Chen,  X. and Sun,  K. and Ren,  J.},
  year = {2018},
  month = may 
}

@article{ndtc-2,
  title = {Strong system-bath coupling induces negative differential thermal conductance and heat amplification in nonequilibrium two-qubit systems},
  volume = {99},
  ISSN = {2470-0053},
  url = {http://dx.doi.org/10.1103/PhysRevE.99.032114},
  number = {3},
  journal = {Physical Review E},
  publisher = {American Physical Society (APS)},
  author = {Liu,  H. and Wang,  C. and Wang,  L. Q. and Ren,  J.},
  year = {2019},
  month = mar 
}

@article{ndtc-3,
  title = {Optimizing the performance of the thermal transistor based on negative differential thermal resistance},
  volume = {124},
  ISSN = {1077-3118},
  url = {http://dx.doi.org/10.1063/5.0201747},
  number = {11},
  journal = {Applied Physics Letters},
  publisher = {AIP Publishing},
  author = {Wu,  T. and Yang,  Y. and Wang,  T. and Li,  X. and Zhang,  L.},
  year = {2024},
  month = mar 
}

@article{ndtc-4,
  title = {Quantum coherence thermal transistors},
  author = {Su,  S. and Zhang,  Y and Andresen,  B. and Chen,  J.},
  journal = {arXiv:1811.02400},
  year = {2018},
 url = {https://arxiv.org/abs/1811.02400}
}

@book{breuer2002theory,
  title={The Theory of Open Quantum Systems},
  author={Breuer, H.P. and Petruccione, F.},
  isbn={9780198520634},
  lccn={2002075713},
  year={2002},
  publisher={Oxford University Press}
}

@article{transistor,
  title = {The Transistor,  A Semi-Conductor Triode},
  volume = {74},
  ISSN = {0031-899X},
  url = {http://dx.doi.org/10.1103/PhysRev.74.230},
  DOI = {10.1103/physrev.74.230},
  number = {2},
  journal = {Physical Review},
  publisher = {American Physical Society (APS)},
  author = {Bardeen,  J. and Brattain,  W. H.},
  year = {1948},
  month = jul,
  pages = {230–231}
}

@article{refrigerator-1,
  title = {Quantum thermodynamic cooling cycle},
  author = {Palao, J. and Kosloff, R. and Gordon, J. M.},
  journal = {Phys. Rev. E},
  volume = {64},
  issue = {5},
  pages = {056130},
  numpages = {8},
  year = {2001},
  month = {Oct},
  publisher = {American Physical Society},
  doi = {10.1103/PhysRevE.64.056130},
  url = {https://link.aps.org/doi/10.1103/PhysRevE.64.056130}
}

@article{refrigerator-2,
  title = {Quantum four-stroke heat engine: Thermodynamic observables in a model with intrinsic friction},
  author = {Feldmann, T. and Kosloff, R.},
  journal = {Phys. Rev. E},
  volume = {68},
  issue = {1},
  pages = {016101},
  numpages = {18},
  year = {2003},
  month = {Jul},
  publisher = {American Physical Society},
  doi = {10.1103/PhysRevE.68.016101},
  url = {https://link.aps.org/doi/10.1103/PhysRevE.68.016101}
}

@article{refrigerator-3,
  title = {How Small Can Thermal Machines Be? The Smallest Possible Refrigerator},
  author = {Linden, N. and Popescu, S. and Skrzypczyk, P.},
  journal = {Phys. Rev. Lett.},
  volume = {105},
  issue = {13},
  pages = {130401},
  numpages = {4},
  year = {2010},
  month = {Sep},
  publisher = {American Physical Society},
  doi = {10.1103/PhysRevLett.105.130401},
  url = {https://link.aps.org/doi/10.1103/PhysRevLett.105.130401}
}

@article{diode,
author = {Yuan, Q. and Wang, T. and Yu, P. and Zhang, H. and Zhang, H.and Ji, W.},
year = {2021},
month = {01},
pages = {106086},
title = {A review on the electroluminescence properties of quantum-dot-lighting-emitting diodes},
volume = {90},
journal = {Organic Electronics},
doi = {10.1016/j.orgel.2021.106086}
}

@article{battery-1,
  title = {Entanglement boost for extractable work from ensembles of quantum batteries},
  author = {Alicki, R. and Fannes, M.},
  journal = {Phys. Rev. E},
  volume = {87},
  issue = {4},
  pages = {042123},
  numpages = {4},
  year = {2013},
  month = {Apr},
  publisher = {American Physical Society},
  doi = {10.1103/PhysRevE.87.042123},
  url = {https://link.aps.org/doi/10.1103/PhysRevE.87.042123}
}

@article{battery-2,
      title={Quantum Batteries - Review Chapter}, 
      author={F. Campaioli and F. A. Pollock and S. Vinjanampathy},
      journal={arxiv:1805.05507},
      year={2018},
     url = {https://arxiv.org/abs/1805.05507}
}

@article{battery-3,
  title = {Quantum thermal machines and batteries},
  author = {Bhattacharjee,  S. and Dutta,  A.},
  journal = {The European Physical Journal B},
  volume = {94},
  ISSN = {1434-6036},
  url = {http://dx.doi.org/10.1140/epjb/s10051-021-00235-3},
  number = {12},
  
  publisher = {Springer Science and Business Media LLC},
  
  year = {2021},
  month = dec 
}

@article{rectifier-1,
  title = {Optimal rectification in the ultrastrong coupling regime},
  author = {Werlang, T. and Marchiori, M. A. and Cornelio, M. F. and Valente, D.},
  journal = {Phys. Rev. E},
  volume = {89},
  issue = {6},
  pages = {062109},
  numpages = {7},
  year = {2014},
  month = {Jun},
  publisher = {American Physical Society},
  doi = {10.1103/PhysRevE.89.062109},
  url = {https://link.aps.org/doi/10.1103/PhysRevE.89.062109}
}

@article{rectifier-2,
  title = {Rectification induced by geometry in two-dimensional quantum spin lattices},
  author = {Chioquetta, A. and Pereira, E. and Landi, Gabriel T. and Drumond, R. C.},
  journal = {Phys. Rev. E},
  volume = {103},
  issue = {3},
  pages = {032108},
  numpages = {6},
  year = {2021},
  month = {Mar},
  publisher = {American Physical Society},
  doi = {10.1103/PhysRevE.103.032108},
  url = {https://link.aps.org/doi/10.1103/PhysRevE.103.032108}
}

@article{rectifier-3,
  title = {Entanglement-enhanced quantum rectification},
  author = {Poulsen, K. and Santos, Alan C. and Kristensen, L. B. and Zinner, N. T.},
  journal = {Phys. Rev. A},
  volume = {105},
  issue = {5},
  pages = {052605},
  numpages = {12},
  year = {2022},
  month = {May},
  publisher = {American Physical Society},
  doi = {10.1103/PhysRevA.105.052605},
  url = {https://link.aps.org/doi/10.1103/PhysRevA.105.052605}
}

@article{sen2023noisyquantumbatteries,
      title={Noisy quantum batteries}, 
      author={Sen,K. and Sen,U.},
    journal = {arXiv:2302.07166},

      year={2023},
      url={https://arxiv.org/abs/2302.07166}, 
}

@article{gksl-ref-1,
    author = {Gorini, V. and Kossakowski, A. and Sudarshan, E. C. G.},
    title = {Completely positive dynamical semigroups of N‐level systems},
    journal = {Journal of Mathematical Physics},
    volume = {17},
    number = {5},
    pages = {821-825},
    year = {1976},
    month = {05},
    issn = {0022-2488},
    doi = {10.1063/1.522979},
    url = {https://doi.org/10.1063/1.522979},

}

@Article{gksl-ref-2,
author={Lindblad, G.},
title={On the generators of quantum dynamical semigroups},
journal={Communications in Mathematical Physics},
year={1976},
month={Jun},
day={01},
volume={48},
number={2},
pages={119-130},
issn={1432-0916},
doi={10.1007/BF01608499},
url={https://doi.org/10.1007/BF01608499}
}

@article{tr1,
  title = {Charge-insensitive qubit design derived from the Cooper pair box},
  author = {Koch, J. and Yu, T. M. and Gambetta, J. and Houck, A. A. and Schuster, D. I. and Majer, J. and Blais, A. and Devoret, M. H. and Girvin, S. M. and Schoelkopf, R. J.},
  journal = {Phys. Rev. A},
  volume = {76},
  issue = {4},
  pages = {042319},
  numpages = {19},
  year = {2007},
  month = {Oct},
  publisher = {American Physical Society},
  doi = {10.1103/PhysRevA.76.042319},
  url = {https://link.aps.org/doi/10.1103/PhysRevA.76.042319}
}

@article{tr2,
title = {Possible new effects in superconductive tunnelling},
journal = {Physics Letters},
volume = {1},
number = {7},
pages = {251-253},
year = {1962},
issn = {0031-9163},
doi = {https://doi.org/10.1016/0031-9163(62)91369-0},
url = {https://www.sciencedirect.com/science/article/pii/0031916362913690},
author = { Josephson, B.D.}
}

@article{tr_exp,
  title = {Magnetic Field Resilience of Three-Dimensional Transmons with Thin-Film ${\text{Al/AlO}}_{x}/\text{Al}$ Josephson Junctions Approaching 1 T},
  author = {Krause, J. and Dickel, C. and Vaal, E. and Vielmetter, M. and Feng, J. and Bounds, R. and Catelani, G. and Fink, J. M. and Ando, Y.},
  journal = {Phys. Rev. Appl.},
  volume = {17},
  issue = {3},
  pages = {034032},
  numpages = {18},
  year = {2022},
  month = {Mar},
  publisher = {American Physical Society},
  doi = {10.1103/PhysRevApplied.17.034032},
  url = {https://link.aps.org/doi/10.1103/PhysRevApplied.17.034032}
}

@article{tr3,
  title = {Optimal quantum resource generation by coupled transmons immersed in Markovian baths},
  author = {Ray, T. and Ghoshal, A. and Rakshit, D. and Sen, U.},
  journal = {Phys. Rev. A},
  volume = {108},
  issue = {5},
  pages = {052417},
  numpages = {12},
  year = {2023},
  month = {Nov},
  publisher = {American Physical Society},
  doi = {10.1103/PhysRevA.108.052417},
  url = {https://link.aps.org/doi/10.1103/PhysRevA.108.052417}
}

@article{apa,
      title={Nonlinearity-assisted advantage for charger-supported open quantum batteries}, 
      author={ Bhattacharyya, A. and  Dongre, P and  Sen, U.},
      year={2024},
 journal = {arXiv:2410.00618},
      url={https://arxiv.org/abs/2410.00618} 
}

@article{kerr1,
      title={Kerr-type nonlinear baths enhance cooling in quantum refrigerators}, 
      author={ Ray, T. and  Mondal, S. and Bhattacharyya, A. and  Ghoshal, A. and Rakshit, D. and  Sen, U.},
      year={2023},
      journal = {arXiv:2311.10499},
      url={https://arxiv.org/abs/2311.10499} 
}

@article{coll-model-1,
  doi = {10.22331/q-2019-06-24-153},
  url = {https://doi.org/10.22331/q-2019-06-24-153},
  title = {Imperfect {T}hermalizations {A}llow for {O}ptimal {T}hermodynamic {P}rocesses},
  author = {B{\"{a}}umer, E. and Perarnau-Llobet, M. and Kammerlander, P. and Wilming, Henrik and Renner, R.},
  journal = {{Quantum}},
  issn = {2521-327X},
  publisher = {{Verein zur F{\"{o}}rderung des Open Access Publizierens in den Quantenwissenschaften}},
  volume = {3},
  pages = {153},
  month = jun,
  year = {2019}
}

@article{coll-model-2,
  title = {Thermalizing Quantum Machines: Dissipation and Entanglement},
  author = {Scarani, V. and Ziman, M. and \ifmmode \check{S}\else \v{S}\fi{}telmachovi\ifmmode \check{c}\else \v{c}\fi{}, P. and Gisin, N. and Bu\ifmmode \check{z}\else \v{z}\fi{}ek, V.},
  journal = {Phys. Rev. Lett.},
  volume = {88},
  issue = {9},
  pages = {097905},
  numpages = {4},
  year = {2002},
  month = {Feb},
  publisher = {American Physical Society},
  doi = {10.1103/PhysRevLett.88.097905},
  url = {https://link.aps.org/doi/10.1103/PhysRevLett.88.097905}
}

@article{coll-model-3,
    author = {Bruneau, L. and Joye, A. and Merkli, M.},
    title = {Repeated interactions in open quantum systems},
    journal = {Journal of Mathematical Physics},
    volume = {55},
    number = {7},
    pages = {075204},
    year = {2014},
    month = {06},
    issn = {0022-2488},
    doi = {10.1063/1.4879240},
    url = {https://doi.org/10.1063/1.4879240},
   
}

@article{coll-model-4,
  title = {Open dynamics under rapid repeated interaction},
  author = {Grimmer, D. and Layden, D. and Mann, Robert B. and Mart\'{\i}n-Mart\'{\i}nez, E.},
  journal = {Phys. Rev. A},
  volume = {94},
  issue = {3},
  pages = {032126},
  numpages = {28},
  year = {2016},
  month = {Sep},
  publisher = {American Physical Society},
  doi = {10.1103/PhysRevA.94.032126},
  url = {https://link.aps.org/doi/10.1103/PhysRevA.94.032126}
}

@article{coll-model-5,
  title = {Quantum and Information Thermodynamics: A Unifying Framework Based on Repeated Interactions},
  author = {Strasberg, P. and Schaller, G. and Brandes, T. and Esposito, M.},
  journal = {Phys. Rev. X},
  volume = {7},
  issue = {2},
  pages = {021003},
  numpages = {33},
  year = {2017},
  month = {Apr},
  publisher = {American Physical Society},
  doi = {10.1103/PhysRevX.7.021003},
  url = {https://link.aps.org/doi/10.1103/PhysRevX.7.021003}
}

@article{coll-model-battery2,
doi = {10.1088/2058-9565/accca4},
url = {https://dx.doi.org/10.1088/2058-9565/accca4},
year = {2023},
month = {may},
publisher = {IOP Publishing},
volume = {8},
number = {3},
pages = {035007},
author = { Morrone, D. and  Rossi, M. A. C. and Smirne,  A. and Genoni, M. G. },
title = {Charging a quantum battery in a non-Markovian environment: a collisional model approach},
journal = {Quantum Science and Technology}
}

@article{coll-model-equi1,
  title = {Nonequilibrium dynamics with finite-time repeated interactions},
  author = {Seah, S. and Nimmrichter, S. and Scarani, V.},
  journal = {Phys. Rev. E},
  volume = {99},
  issue = {4},
  pages = {042103},
  numpages = {11},
  year = {2019},
  month = {Apr},
  publisher = {American Physical Society},
  doi = {10.1103/PhysRevE.99.042103},
  url = {https://link.aps.org/doi/10.1103/PhysRevE.99.042103}
}

@article{coll-model-equi2,
  title = {Collision Models Can Efficiently Simulate Any Multipartite Markovian Quantum Dynamics},
  author = {Cattaneo, M. and De Chiara, G. and Maniscalco, S. and Zambrini, R. and Giorgi, G. L.},
  journal = {Phys. Rev. Lett.},
  volume = {126},
  issue = {13},
  pages = {130403},
  numpages = {8},
  year = {2021},
  month = {Apr},
  publisher = {American Physical Society},
  doi = {10.1103/PhysRevLett.126.130403},
  url = {https://link.aps.org/doi/10.1103/PhysRevLett.126.130403}
}

@article{coll-model-str-coupling,
  title = {Repeated Interactions and Quantum Stochastic Thermodynamics at Strong Coupling},
  author = {Strasberg, P.},
  journal = {Phys. Rev. Lett.},
  volume = {123},
  issue = {18},
  pages = {180604},
  numpages = {7},
  year = {2019},
  month = {Oct},
  publisher = {American Physical Society},
  doi = {10.1103/PhysRevLett.123.180604},
  url = {https://link.aps.org/doi/10.1103/PhysRevLett.123.180604}
}

@article{coll-model-thermo1,
  title = {Thermodynamics of Weakly Coherent Collisional Models},
  author = {Rodrigues, F. L. S. and De Chiara, G. and Paternostro, M. and Landi, G. T.},
  journal = {Phys. Rev. Lett.},
  volume = {123},
  issue = {14},
  pages = {140601},
  numpages = {6},
  year = {2019},
  month = {Oct},
  publisher = {American Physical Society},
  doi = {10.1103/PhysRevLett.123.140601},
  url = {https://link.aps.org/doi/10.1103/PhysRevLett.123.140601}
}

@article{coll-model-thermo2,
doi = {10.1088/1367-2630/abeb47},
url = {https://dx.doi.org/10.1088/1367-2630/abeb47},
year = {2021},
month = {apr},
publisher = {IOP Publishing},
volume = {23},
number = {4},
pages = {043024},
author = { Hammam, K. and Hassouni, Y. and  Fazio, R. and Manzano, G.},
title = {Optimizing autonomous thermal machines powered by energetic coherence},
journal = {New Journal of Physics}
}

@book{trans-book-1,
  title={Electronic Devices and Circuits},
  author={Millman, J. and Halkias, C.C.},
  isbn={9780070423800},
  lccn={lc67016934},
  series={Electrical Engineering Series},
  year={1967},
  publisher={McGraw-Hill}
}

@book{trans-book-2,
  title={Integrated Electronics: Analog and Digital Circuits and Systems},
  author={Millman, J. and Halkias, C.C.},
  isbn={9780070423152},
  lccn={lc79172657},
  series={Electrical Engineering Series},
  year={1972},
  publisher={McGraw-Hill}
}

@book{trans-book-3,
  title={Electronic Devices and Circuit Theory},
  author={Boylestad, R.L. and Nashelsky, L.},
  isbn={9789702604365},
  lccn={2001021973},
  year={2002},
  publisher={Prentice Hall}
}

@article{trans-work-old2,
  title = {Physical Principles Involved in Transistor Action},
  author = {Bardeen, J. and Brattain, W. H.},
  journal = {Phys. Rev.},
  volume = {75},
  issue = {8},
  pages = {1208--1225},
  numpages = {0},
  year = {1949},
  month = {Apr},
  publisher = {American Physical Society},
  doi = {10.1103/PhysRev.75.1208},
  url = {https://link.aps.org/doi/10.1103/PhysRev.75.1208}
}

@ARTICLE{trans-work-old3,
  author={Brinkman, W.F. and Haggan, D.E. and Troutman, W.W.},
  journal={IEEE Journal of Solid-State Circuits}, 
  title={A history of the invention of the transistor and where it will lead us}, 
  year={1997},
  volume={32},
  number={12},
  pages={1858-1865},
  doi={10.1109/4.643644}}

@article{breuer_,
  title = {Correlated projection operator approach to non-Markovian dynamics in spin baths},
  author = {Fischer, J. and Breuer, H.},
  journal = {Phys. Rev. A},
  volume = {76},
  issue = {5},
  pages = {052119},
  numpages = {10},
  year = {2007},
  month = {Nov},
  publisher = {American Physical Society},
  doi = {10.1103/PhysRevA.76.052119},
  url = {https://link.aps.org/doi/10.1103/PhysRevA.76.052119}
}

@article{samyadeb,
  title = {Exact master equation for a spin interacting with a spin bath: Non-Markovianity and negative entropy production rate},
  author = {Bhattacharya, S. and Misra, A. and Mukhopadhyay, C. and Pati, A. K.},
  journal = {Phys. Rev. A},
  volume = {95},
  issue = {1},
  pages = {012122},
  numpages = {10},
  year = {2017},
  month = {Jan},
  publisher = {American Physical Society},
  doi = {10.1103/PhysRevA.95.012122},
  url = {https://link.aps.org/doi/10.1103/PhysRevA.95.012122}
}

@article{apa_spin_ref,
  title = {Transient effects in quantum refrigerators with finite environments},
  author = {Bhattacharyya, A. and Ghoshal, A. and Sen, U.},
  journal = {Phys. Rev. A},
  volume = {111},
  issue = {1},
  pages = {012209},
  numpages = {18},
  year = {2025},
  month = {Jan},
  publisher = {American Physical Society},
  doi = {10.1103/PhysRevA.111.012209},
  url = {https://link.aps.org/doi/10.1103/PhysRevA.111.012209}
}

@article{arma1,
author = {Sanderson, C. and Curtin, R.},
month = {07},
pages = {26},
title = {Armadillo: A template-based C++ library for linear algebra},
volume = {1},
journal = {Journal of Open Source Software},
year = {2016},
doi = {10.21105/joss.00026}
}

@article{arma2,
author="Sanderson, C.
and Curtin, R.",
editor="Davenport, J. H.
and Kauers, M.
and Labahn, G.
and Urban, J.",
title="A User-Friendly Hybrid Sparse Matrix Class in C++",
booktitle="Mathematical Software -- ICMS 2018",
year="2018",
publisher="Springer International Publishing",
address="Cham",
pages="422--430",
isbn="978-3-319-96418-8"
}

@misc{Fq_transistor_2,
      title={Fluctuations and optimal control in a Floquet Quantum Thermal Transistor}, 
      author={ Das, S. and  Mahunta, S. and Gupt, N. and  Mukherjee, V. and Ghosh, A.},
      journal={arXiv:2412.16920},
      primaryClass={quant-ph},
      url={https://arxiv.org/abs/2412.16920}, 
}

@article{c_1,
title = {Quantum collision models: Open system dynamics from repeated interactions},
journal = {Physics Reports},
volume = {954},
pages = {1-70},
year = {2022},
note = {Quantum collision models: Open system dynamics from repeated interactions},
issn = {0370-1573},
doi = {https://doi.org/10.1016/j.physrep.2022.01.001},
url = {https://www.sciencedirect.com/science/article/pii/S0370157322000035},
author = {Ciccarello, F. and  Lorenzo, S. and Giovannetti, V. and Palma, G. M.}
}

@article{c_2,
author = {Cattaneo, M. and Giorgi, G. L. and Zambrini, R. and Maniscalco, S.},
title = {A Brief Journey through Collision Models for Multipartite Open Quantum Dynamics},
journal = {Open Systems \& Information Dynamics},
volume = {29},
number = {03},
pages = {2250015},
year = {2022},
doi = {10.1142/S1230161222500159},

URL = { 
    
        https://doi.org/10.1142/S1230161222500159
    
    

},
eprint = { 
    
        https://doi.org/10.1142/S1230161222500159
    
    

}
}

@article{cc3,
doi = {10.1088/1367-2630/abd57d},
url = {https://doi.org/10.1088/1367-2630/abd57d},
year = {2021},
month = {mar},
publisher = {IOP Publishing},
volume = {23},
number = {3},
pages = {033031},
author = {Chisholm, D. A. and García-Pérez, G. and Rossi, M. A. C. and Palma, G. M. and Maniscalco, S.},
title = {Stochastic collision model approach to transport phenomena in quantum networks},
journal = {New Journal of Physics}
}

@article{C_4,
doi = {10.1088/2058-9565/aded2e},
url = {https://doi.org/10.1088/2058-9565/aded2e},
year = {2025},
month = {jul},
publisher = {IOP Publishing},
volume = {10},
number = {3},
pages = {035066},
author = {Pezzutto, M. and De Chiara, G. and Gherardini, S.},
title = {Non-positive energy quasidistributions in coherent collision models},
journal = {Quantum Science and Technology}
}

@article{C_5,
doi = {10.1088/1367-2630/ad202a},
url = {https://doi.org/10.1088/1367-2630/ad202a},
year = {2024},
month = {feb},
publisher = {IOP Publishing},
volume = {26},
number = {2},
pages = {023001},
author = {Cusumano, Stefano and De Chiara, Gabriele},
title = {Structured quantum collision models: generating coherence with thermal resources},
journal = {N. J. P}
}

@article{Camati2020,
  author  = {Camati, P. A. and Santos, J. F. G. and Serra, R. M.},
  title   = {Employing non-Markovian effects to improve the performance of a quantum Otto refrigerator},
  journal = {Phys. Rev. A},
  volume  = {102},
  number  = {1},
  pages   = {012217},
  year    = {2020},
  doi     = {10.1103/PhysRevA.102.012217}
}

@article{Uzdin2016,
  author  = {Uzdin, R. and Levy, A. and Kosloff, R.},
  title   = {Quantum heat machines equivalence and work extraction beyond Markovianity, and strong coupling via heat exchangers},
  journal = {Entropy},
  volume  = {18},
  number  = {4},
  pages   = {124},
  year    = {2016},
  doi     = {10.3390/e18040124}
}

@article{Kato2016,
  author  = {Kato, A. and Tanimura, Y.},
  title   = {Quantum heat current under non-perturbative and non-Markovian conditions: Applications to heat machines},
  journal = {J. Chem. Phys.},
  volume  = {145},
  number  = {22},
  pages   = {224105},
  year    = {2016},
  doi     = {10.1063/1.4971370}
}

@article{Chen2016,
  author  = {Chen, H.-B. and Chiu, P.-Y. and Chen, Y.-N.},
  title   = {Vibration-induced coherence enhancement of the performance of a biological quantum heat engine},
  journal = {Phys. Rev. E},
  volume  = {94},
  number  = {5},
  pages   = {052101},
  year    = {2016},
  doi     = {10.1103/PhysRevE.94.052101}
}

@article{DasMukherjee2020,
  author  = {Das, A. and Mukherjee, V.},
  title   = {A quantum enhanced finite-time Otto cycle},
  journal = {Phys. Rev. Research},
  volume  = {2},
  number  = {3},
  pages   = {033083},
  year    = {2020},
  doi     = {10.1103/PhysRevResearch.2.033083}
}

@article{Shirai2021,
  author  = {Shirai, Y. and Hashimoto, K. and Tezuka, R. and Uchiyama, C. and Hatano, N.},
  title   = {Non-Markovian effect on quantum Otto engine: Role of system--reservoir interaction},
  journal = {Phys. Rev. Research},
  volume  = {3},
  number  = {2},
  pages   = {023078},
  year    = {2021},
  doi     = {10.1103/PhysRevResearch.3.023078}
}

@article{Raja2021,
  author  = {Raja, S. H. and Maniscalco, S. and Paraoanu, G.-S. and Pekola, J. P. and Lo Gullo, N.},
  title   = {Finite-time quantum Stirling heat engine},
  journal = {New J. Phys.},
  volume  = {23},
  number  = {3},
  pages   = {033034},
  year    = {2021},
  doi     = {10.1088/1367-2630/abeb7d}
}

@article{Chakraborty2022,
  author  = {Chakraborty, S. and Das, A. and Chru\'sci\'nski, D.},
  title   = {Strongly coupled quantum Otto cycle with single qubit bath},
  journal = {Phys. Rev. E},
  volume  = {106},
  number  = {6},
  pages   = {064133},
  year    = {2022},
  doi     = {10.1103/PhysRevE.106.064133}
}

@article{Cavaliere2022,
  author  = {Cavaliere, F. and Carrega, M. and De Filippis, G. and Cataudella, V. and Benenti, G. and Sassetti, M.},
  title   = {Dynamical heat engines with non-Markovian reservoirs},
  journal = {Phys. Rev. Research},
  volume  = {4},
  number  = {3},
  pages   = {033233},
  year    = {2022},
  doi     = {10.1103/PhysRevResearch.4.033233}
}

@article{Ptaszynski2022,
  author  = {Ptaszy\'nski, K.},
  title   = {Non-Markovian thermal operations boosting the performance of quantum heat engines},
  journal = {Phys. Rev. E},
  volume  = {106},
  number  = {1},
  pages   = {014114},
  year    = {2022},
  doi     = {10.1103/PhysRevE.106.014114}
}

@article{PhysRevB.70.045323,
  title = {Non-Markovian dynamics in a spin star system: Exact solution and approximation techniques},
  author = {Breuer, Heinz-Peter and Burgarth, Daniel and Petruccione, Francesco},
  journal = {Phys. Rev. B},
  volume = {70},
  issue = {4},
  pages = {045323},
  numpages = {10},
  year = {2004},
  month = {Jul},
  publisher = {American Physical Society},
  doi = {10.1103/PhysRevB.70.045323},
  url = {https://link.aps.org/doi/10.1103/PhysRevB.70.045323}
}

@article{PhysRevB.104.245418,
  title = {Entanglement and work extraction in the central-spin quantum battery},
  author = {Liu, J. and Shi, H. and Shi, Y. and Wang, X. and Yang, W.},
  journal = {Phys. Rev. B},
  volume = {104},
  issue = {24},
  pages = {245418},
  numpages = {8},
  year = {2021},
  month = {Dec},
  publisher = {American Physical Society},
  doi = {10.1103/PhysRevB.104.245418},
  url = {https://link.aps.org/doi/10.1103/PhysRevB.104.245418}
}

@article{PhysRevA.81.062353,
  title = {Channel capacities of an exactly solvable spin-star system},
  author = {Arshed, N. and Toor, A. H. and Lidar, D. A.},
  journal = {Phys. Rev. A},
  volume = {81},
  issue = {6},
  pages = {062353},
  numpages = {8},
  year = {2010},
  month = {Jun},
  publisher = {American Physical Society},
  doi = {10.1103/PhysRevA.81.062353},
  url = {https://link.aps.org/doi/10.1103/PhysRevA.81.062353}
}

@article{4hr3-rl2y,
  title = {Optimal performances of a quantum battery via non-Markovian bath modulation},
  author = {Li, Y. and Liu, R. F. and You, J. B. and Yang, W. L. and Guan, H.},
  journal = {Phys. Rev. E},
  volume = {112},
  issue = {6},
  pages = {064106},
  numpages = {11},
  year = {2025},
  month = {Dec},
  publisher = {American Physical Society},
  doi = {10.1103/4hr3-rl2y},
  url = {https://link.aps.org/doi/10.1103/4hr3-rl2y}
}

@article{Vacchini2018NonMarkovianThermo,
  author  = {Vacchini, B. and Smirne, A. and Laine, E. and Piilo, J. and Breuer, H.P},
  title   = {Thermodynamics of non-Markovian reservoirs and heat engines},
  journal = {Phys. Rev. E},
  volume  = {97},
  pages   = {062108},
  year    = {2018},
  doi     = {10.1103/PhysRevE.97.062108}
}

@article{KosloffLevy2014,
  author  = {Kosloff, R. and Levy, A.},
  title   = {Quantum Heat Engines and Refrigerators: Continuous Devices},
  journal = {Annual Review of Physical Chemistry},
  volume  = {65},
  pages   = {365--393},
  year    = {2014},
  doi     = {10.1146/annurev-physchem-040513-103724}
}

@article{UzdinLevyKosloff2015,
  author  = {Uzdin, R. and Levy, A. and Kosloff, R.},
  title   = {Equivalence of Quantum Heat Machines, and Quantum-Thermodynamic Signatures},
  journal = {Phys. Rev. X},
  volume  = {5},
  pages   = {031044},
  year    = {2015},
  doi     = {10.1103/PhysRevX.5.031044}
}

@article{Saha2023Harnessing,
  author       = {Saha, D. and Bhattacharyya, A. and Sen, K. and Sen, U.},
  title        = {Harnessing energy extracted from heat engines to charge quantum batteries},
  journal      = {arXiv:2309.15634},
  year         = {2023},
  eprint       = {2309.15634}
}

@incollection{NimmrichterRouletScarani2018,
  author    = {Nimmrichter, S. and Roulet, A. and Scarani, V.},
  title     = {Quantum Rotor Engines},
  booktitle = {Thermodynamics in the Quantum Regime: Fundamental Aspects and New Directions},
  editor    = {Binder, F. and Correa, L. A. and Gogolin, C. and Anders, J. and Adesso, G.},
  publisher = {Springer International Publishing},
  address   = {Cham},
  pages     = {227--245},
  year      = {2018},
  doi       = {10.1007/978-3-319-99046-0_10}
}

@article{PhysRevResearch.5.023066,
  title = {Switching the function of the quantum Otto cycle in non-Markovian dynamics: Heat engine, heater, and heat pump},
  author = {Ishizaki, M. and Hatano, N. and Tajima, H.},
  journal = {Phys. Rev. Res.},
  volume = {5},
  issue = {2},
  pages = {023066},
  numpages = {14},
  year = {2023},
  month = {Apr},
  publisher = {American Physical Society},
  doi = {10.1103/PhysRevResearch.5.023066},
  url = {https://link.aps.org/doi/10.1103/PhysRevResearch.5.023066}
}

@article{
Avijit_Mishra,
author = {Tomáš Opatrný  and Šimon Bräuer  and Abraham G. Kofman  and Avijit Misra  and Nilakantha Meher  and Ofer Firstenberg  and Eilon Poem  and Gershon Kurizki },
title = {Nonlinear coherent heat machines},
journal = {Science Advances},
volume = {9},
number = {1},
pages = {eadf1070},
year = {2023},
doi = {10.1126/sciadv.adf1070}}

@article{BLP,
  title = {Measure for the Degree of Non-Markovian Behavior of Quantum Processes in Open Systems},
  author = {Breuer, H. P. and Laine, E. M. and Piilo, J.},
  journal = {Phys. Rev. Lett.},
  volume = {103},
  issue = {21},
  pages = {210401},
  numpages = {4},
  year = {2009},
  month = {Nov},
  publisher = {American Physical Society},
  doi = {10.1103/PhysRevLett.103.210401},
  url = {https://link.aps.org/doi/10.1103/PhysRevLett.103.210401}
}

@article{Transiant_transistor,
  title = {Quantum thermal transistors: Operation characteristics in steady state versus transient regimes},
  author = {Ghosh, R. and Ghoshal, A. and Sen, U.},
  journal = {Phys. Rev. A},
  volume = {103},
  issue = {5},
  pages = {052613},
  numpages = {14},
  year = {2021},
  month = {May},
  publisher = {American Physical Society},
  doi = {10.1103/PhysRevA.103.052613},
  url = {https://link.aps.org/doi/10.1103/PhysRevA.103.052613}
}

\end{document}